\documentclass[aps,pra,10pt,english,showpacs,floatfix,twocolumn,twoside]{revtex4-1}

\usepackage{amsmath,amstext,amssymb,babel,graphicx,geometry,bm,dcolumn,color}
\usepackage[colorlinks=true,citecolor=blue,urlcolor=red]{hyperref}
\usepackage[table]{xcolor}
\begin{document}

\title{Reducing Computational Complexity of Quantum Correlations
}
\author{
Titas Chanda, Tamoghna Das, Debasis Sadhukhan, Amit Kumar Pal, Aditi Sen(De), and Ujjwal Sen
}
\affiliation{Harish-Chandra Research Institute, Chhatnag Road, Jhunsi, Allahabad - 211019, India}

\begin{abstract}
  We address the issue of reducing the resource required to compute information-theoretic quantum correlation measures 
  like quantum 
  discord and quantum work deficit in two qubits and higher dimensional systems. 
  We show that determination of 
  the quantum correlation measure 
  is possible even if we utilize a restricted set of local measurements. 
  We find that the determination allows us to obtain a closed form of quantum discord and quantum work 
  deficit for several classes of states, with a low error. We 
  show that the computational error caused by the constraint over the complete set of local measurements 
  reduces fast with an increase in the size of the restricted set, 
  implying usefulness of constrained optimization, especially 
  with the increase of dimensions. We perform quantitative analysis to investigate how 
  the error scales with the system size, taking into account a set of plausible constructions of 
  the constrained set. Carrying out 
  a comparative study, we show that the resource required to optimize 
  quantum work deficit is usually higher than that required for 
  quantum discord. We also demonstrate that minimization of quantum discord and quantum work deficit is easier in the case of 
  two-qubit mixed states of fixed ranks and with positive partial transpose in comparison to the corresponding states
  having non-positive partial transpose. Applying the 
  methodology to quantum spin models, 
  we show that the constrained optimization can be used with advantage in analyzing such systems in quantum 
  information-theoretic language. For bound entangled states, we show that the 
  error is significantly low when the measurements 
  correspond to the spin observables along the three Cartesian coordinates, and thereby we obtain 
  expressions of quantum discord and quantum work deficit for these bound entangled states.    
\end{abstract}


\maketitle

\section{Introduction}
\label{intro}

Entanglement \cite{ent_horodecki} as a measure of quantum correlations existing between 
subsystems of a composite quantum system has been shown to be indispensable in performing several
quantum information tasks \cite{ent_th,ent_exp}. To deal with challenges such as decoherence due to 
system-environment interaction, entanglement distillation protocols \cite{distil} to purify highly entangled states from a 
collection of states with relatively low entanglement have also been invented. 
Parallely, various counter-intuitive findings such as substantial non-classical 
efficiency of quantum states with vanishingly small entanglement, and locally indistinguishable orthogonal 
product states \cite{nl_w_ent,opt_det,dqci,dakic_geom}
have motivated the search for quantum correlations not belonging to the entanglement-separability paradigm. 
This has led to the possibility of introducing more fine-grained quantum correlation measures than entanglement, such as 
quantum discord (QD) \cite{disc_group}, quantum work deficit (QWD) \cite{wdef_group}, and various 
\textquoteleft discord-like\textquoteright\, measures \cite{modi_rmp_2012,celeri_2011}, opening up a new direction of 
research in quantum 
information theory. Although establishing a link between the measures of quantum correlations belonging to 
the two different genres has also 
been tried \cite{ent_disc}, a decisive result is yet to be found in the case of mixed bipartite quantum states. Note, however, 
that all these measures reduce to von Neumann entropy of local density matrix for pure states. 
In recent years, the interplay between entanglement distillation
and quantum correlations such as QD and QWD has been under focus \cite{distil_disc}. However, proper 
understanding of the relation between 
such measures and distillable as well as bound entanglement \cite{be1,be2,be3,be4,be5,be6,npptbe} is yet to be achieved.

There has been a substantial amount of work in determining QD for various classes of bipartite as well as multipartite 
mixed quantum states \cite{luo_pra_2008,disc_2q,disc_mq_hd}. 
A common observation that stands out from these works is the computational complexity of the task,  
due to the optimization over a complete set of local measurements involved in its definition \cite{disc_group}. 
For general quantum states, the optimization is often achieved via numerical techniques.
It has recently been shown that the problem of 
computing quantum discord is NP-complete, thereby making the quantity computationally intractable \cite{huang_disc_np}.
The lack of a well-established analytic treatment to determine QD has also restricted the number of 
experiments in this topic \cite{disc_expt}.  
Despite considerable efforts to analytically determine QD for general two-qubit states \cite{luo_pra_2008,disc_2q}, 
a closed form expression exists only for the Bell diagonal (BD) states \cite{luo_pra_2008}.

A number of recent numerical studies have shown that for a large fraction of a very special class of two-qubit states,
QD can always be calculated by performing the optimization over only a small subset of the complete set of local projection 
measurements \cite{num-err_group,freezing_paper}, thereby reducing the computational difficulty to a great extent. 
These states are constructed of the three diagonal correlators of the correlation matrix, and any one of the three magnetizations,  
which is similar for both the qubits. The subset, in the present case, consists of the projection measurements corresponding 
to the three Pauli matrices, $\sigma^{x},\sigma^{y}$, and $\sigma^{z}$. 
Curiously, the assumption that QD can be optimized over this subset for the entire class of 
such states results only in a small absolute error in the case of those states where the assumption is 
not valid \cite{num-err_group,freezing_paper}. This property also allows one to determine closed 
form for QD for the entire class of such two-qubit states within the margin of small absolute error in 
calculation \cite{num-err_group,freezing_paper,disc_num_anal}.

Optimization over such a small subset of the complete set of local projection measurements
is logical in the situations where a constraint over the allowed local measurements is at work. 
The knowledge of such a subset may be important in quantum estimation theory \cite{qet}, 
in relation to quantities that are non-linear functions of the quantum states, such as the QD, where   
estimation of the parameter values to determine the optimal projection measurement depends on the 
the number of measurements required to obtain the desired quantity within a manageable range of error. Also, the existence of 
such subsets has the potential to be  operationally as well as energetically advantageous in experimental determination of the QD and 
similar measures for a given quantum state. However, the investigation of the existence of such \textit{special} subsets of allowed 
local projectors in the computation of quantum correlations like QD and QWD, as of now, are confined only to special classes of 
quantum states in $\mathbb{C}^{2}\otimes\mathbb{C}^{2}$ systems \cite{num-err_group,freezing_paper,disc_num_anal}.
A natural question that arises is whether such subsets of local projectors can exist for general 
bipartite quantum states. The possible scaling of the absolute error resulting from the limitation on the number of 
allowed local projectors in the subset is also an interesting issue.

In this paper, we investigate the issue of simplifying the optimization of quantum correlation measures like 
QD and QWD in the case of general two-qubit mixed states of different ranks with positive as well as non-positive 
partial transpose (NPPT) \cite{ppt-ph} by using a restricted set of local projection measurements. 
We provide a mathematical description of the optimization of quantum correlation measures 
over a restricted set of allowed local projection measurements, and discuss the related statistics of computational error. 
Using a set of plausible definitions of the restricted set, 
we show that the absolute error, resulting due to the constraint over the set 
of projectors, dies out considerably fast. Using the scaling of the error with the size of the restricted 
subset, we demonstrate that even a small number of properly chosen projectors can form a restricted set leading to a very 
small absolute error, thereby making the computation of quantum correlation measures considerably easier. 

Our method also helps us to find 
expressions with negligible error for QD and QWD of several important classes of quantum states. 
We demonstrate this in the case of two-qubit ``$X$'' states \cite{xstate}, which occur, in general, in
ground or thermal states of several quantum spin models.  
Hence, our approach provides a way to study cooperative phenomena present
in such systems with less numerical difficulty, as demonstrated here for anisotropic $XY$ model in the 
presence of external transverse field \cite{xy_group}.
We extend the study 
to the paradigmatic classes of bound entangled (BE) states, where the computation of QD and QWD using a special restricted subset is discussed,
and their analytic forms with small error are determined. 
The results indicate that the error resulting from the restricted measurement in the case of states with positive partial transpose 
(PPT) is less compared to the states with non-positive partial transpose (NPPT). 
Such constrained optimizations can be a powerful 
tool to study physical quantities in higher dimensions, and to obtain closed forms of QD and QWD.

The paper is organized as follows. In Sec. \ref{definitions}, we provide a mathematical description of the computation of quantum 
correlation measures by performing the optimization over a constrained set of local projection measurements, defining 
the corresponding absolute error in calculation. In Sec. \ref{twoqubit}, we discuss a set of constructions of the subset in 
relation to the statistics of the error for general two-qubit mixed states in the state space. 
We also consider a general two-qubit state in the parameter space, and show how symmetry of the state helps in defining the 
restricted subset. We demonstrate how our method can be used to determine closed form expressions 
of quantum correlation measures, and comment on the applicability of the method in real physical systems like quantum 
spin models. In particular, we show that the constrained optimization technique performs quite well in analyzing the 
anisotropic XY model in transverse field in terms of quantum correlation measures. 
In Sec. \ref{bestates}, the results on the quantum correlations in  BE states are presented. 
Sec. \ref{conclude} contains the concluding remarks.

\section{Definitions and methodology}
\label{definitions}

In this section, after presenting an overview of the quantum correlation measures used in this paper, namely, QD and QWD, we 
introduce the main concepts of the paper, i.e., constrained QD as well as QWD by  
restricting the optimization over a small subset of the complete set of allowed local measurements. We also discuss the corresponding 
error generated due to the limitation in measurements.

\subsection{Quantum discord}
\label{def_qd}

For a bipartite quantum state $\rho_{AB}$, the QD is defined as the minimum difference between two inequivalent definitions of the 
quantum mutual information. While one of them, given by 
$I(\rho_{AB})=S\left(\rho_{A}\right)+S\left(\rho_{B}\right)-S\left(\rho_{AB}\right)$,
can be identified as the \textquotedblleft total correlation\textquotedblright\; of the 
bipartite quantum system $\rho_{AB}$ \cite{tot_corr}, 
the other definition takes the form $J_{\rightarrow}(\rho_{AB})=S\left(\rho_{B}\right)-S\left(\rho_{B}|\rho_{A}\right)$,
which can be argued as a measure of classical correlation \cite{disc_group}. 
Here, $\rho_{A}$ and $\rho_{B}$ are local density matrices of the subsystems $A$ and $B$, respectively, 
$S\left(\rho\right)=-\mbox{Tr}\left[\rho\log_{2}\rho\right]$ is the von Neumann entropy of the quantum state $\rho$, 
and $S(\rho_{B}|\rho_{A})=\sum_{k}p_{k}S\left(\rho_{AB}^{k}\right)$ is the quantum conditional entropy with 
\begin{eqnarray}
\rho_{AB}^{k}=\left(\Pi_{k}^{A}\otimes I_{B}\right)\rho_{AB}\left(\Pi_{k}^{A}\otimes I_{B}\right)/p_{k},
\label{aftermeasure}
\end{eqnarray}
and 
\begin{eqnarray}
p_{k}=\mbox{Tr}\left[\left(\Pi_{k}^{A}\otimes I_{B}\right)\rho_{AB}\right]. 
\label{probnorm}
\end{eqnarray}
The subscript `$\rightarrow$' implies that the measurement, represented by a complete set of rank-$1$ projective operators, 
$\{ \Pi_{k}^{A} \}$, 
is performed locally on the subsystem $A$, and $I_{B}$ is the identity operator defined over the Hilbert space of the subsystem $B$. 
The QD is thus quantified as $D=\min\{I(\rho_{AB})-J_{\rightarrow}(\rho_{AB})\}$, where the minimization is performed over
the set $\mathcal{S}_{C}$, the class of all complete sets of rank-$1$ projective operators. One must note here the asymmetry embedded in the definition of the QD over the 
interchange of the two subsystems, $A$ and $B$. Throughout this paper, we calculate QD by performing 
local measurement on the subsystem $A$.

\subsection{Quantum work deficit}
\label{def_qwd}

Along with the QD, we also consider the QWD~\cite{wdef_group} of a quantum state, defined 
as the difference between the amount of extractable pure states 
under suitably restricted global and local operations. In the case of a bipartite state $\rho_{AB}$, the class of global operations, 
consisting of \textit{(i)} unitary operations, and \textit{(ii)} dephasing the bipartite state by a
set of projectors, $\{\Pi_{k}\}$, defined  on the Hilbert space $\mathcal{H}$ of $\rho_{AB}$, is called 
\textquotedblleft closed operations\textquotedblright\; (CO) under which the amount of extractable pure states from $\rho_{AB}$ is given by 
$I_{\mbox{\scriptsize CO\normalsize}}=\log_{2}\mbox{dim}\left(\mathcal{H}\right)-S(\rho_{AB})$. And, the class of operations 
consisting of \textit{(i)} local unitary operations,  \textit{(ii)} 
dephasing by local measurement on the subsystem $A$, and \textit{(iii)} communicating the dephased 
subsystem to the other party, $B$, over a noiseless quantum channel is the class of \textquotedblleft closed local operations and 
classical communication\textquotedblright\; (CLOCC), under which the extractable amount of pure states is   
$I_{\mbox{\scriptsize CLOCC\normalsize}}=\log_{2}\mbox{dim}\left(\mathcal{H}\right)-\min
 S\left(\rho_{AB}^{\prime}\right)$. Here, $\rho_{AB}^{\prime}=\sum_{k} p_{k}\rho_{AB}^{k}$ is the average quantum state after 
the projective measurement $\{\Pi_{k}^{A}\}$ has been performed on $A$, with $\rho_{AB}^{k}$ and $p_{k}$ given by 
Eqs. (\ref{aftermeasure}) and (\ref{probnorm}), respectively. The minimization in $I_{\mbox{\scriptsize CLOCC\normalsize}}$ 
is achieved over $\mathcal{S}_{C}$. The QWD, $W$, is given by  the difference between the quantities $I_{\mbox{\scriptsize CO\normalsize}}\left(\rho_{AB}\right)$
and $I_{\mbox{\scriptsize CLOCC\normalsize}}\left(\rho_{AB}\right)$.

\subsection{Constrained Quantum Correlations: Error in Estimation} 
\label{def_err}

We now introduce the physical quantity, which will help us to reduce the computational complexity involved in evaluation of QD and QWD.  
In particular, we consider the bipartite quantum correlations in the scenario where there are restrictions on the complete set 
of projectors defining the local measurement on 
one of the subsystems. Let us assume that constraints on the local measurement restrict the class of projection measurements to a subset 
$\mathcal{S}_{E}$ $(\mathcal{S}_{E}\subseteq\mathcal{S}_{C})$, where there are $n$ sets of projection measurements in $\mathcal{S}_E$. 
Performing the optimization only over 
the set $\mathcal{S}_{E}$, a \textquotedblleft constrained\textquotedblright\, quantum 
correlation (CQC), $Q_{c}$, can be defined. We call the subset $\mathcal{S}_{E}$ as the \textquotedblleft earmarked\textquotedblright\, 
set. 
Let the actual value of a given quantum correlation measure, $Q$, for a fixed bipartite state, $\rho_{AB}$, be $Q_{a}$. 
If the definition of $Q$ involves a minimization, $\log_2d\geq Q_{c}\geq Q_{a}$, while a maximization in the definition
leads to $\log_2d\geq Q_{a}\geq Q_{c}$, where $d$ is the minimum of the dimensions among the two parties making up the bipartite state 
and where we have assumed that $\log_2d$ is the maximum value of $Q_a$ or $Q_c$. 
For example, one can define the constrained QD (CQD) as  
\begin{eqnarray}
 D_{c}=\underset{\mathcal{S}_{E}^{D}}{\min}[I(\rho_{AB})-J_{\rightarrow}(\rho_{AB})],
\end{eqnarray}
while the constrained QWD (CQWD) is given by
\begin{eqnarray}
 W_{c}=\underset{\mathcal{S}_{E}^{W}}{\min}[S(\rho_{AB}^{\prime})-S(\rho_{AB})].
\end{eqnarray} 
Note that in general, the earmarked sets for QD and QWD, represented by $\mathcal{S}_{E}^{D}$ and 
$\mathcal{S}_{E}^{W}$ respectively, may not be identical.

Evidently, the actual projector for which the quantum correlation is optimized may not belong to $\mathcal{S}_E$. Therefore,  
restricting the optimization over the earmarked set gives rise to error in estimation of the value of $Q$ for a fixed
quantum state, $\rho_{AB}$.   
Let us denote the \emph{absolute} error occurring due to the optimization over $S_{E}$, instead of $S_{C}$, for an arbitrary bipartite 
state $\rho_{AB}$, by $\varepsilon$, where 
\begin{eqnarray}
 \varepsilon_n=|Q_{c}-Q_{a}|,
 \label{error}
\end{eqnarray}
with  $\log_2d\geq\varepsilon_n\geq0$. We call this error as the ``voluntary'' error 
(VE). One must note that the VE depends on the size of the earmarked set, $n$, as well as the distribution of the elements of 
$\mathcal{S}_{E}$ in the space of projection measurements. 
When $n\rightarrow\infty$, we may (but not necessarily) have $\mathcal{S}_E\rightarrow\mathcal{S}_C$, resulting in 
$Q_{c}\rightarrow Q_{a}$, whence VE vanishes. 
However, for a finite value of $n$, we denote the VE 
by $\varepsilon_{n}$. Note that $\varepsilon_n$ also depends on the actual form of the $n$ projection measurements in $\mathcal{S}_E$. 
If $\varepsilon=0$ for a quantum state even when the optimization is performed over the set $\mathcal{S}_{E}$, 
we call the quantum state an ``exceptional'' state.

% We conclude this section by pointing out that the VE for QD is unity only in the rare case of zero-discordant states 
% with $D_{c}=1$. Similar result is expected for QWD also.  

Apart from the quantum information-theoretic measures having entropic definitions, 
such as QD and QWD, there exists a 
variety of geometric measures of ``discord-like'' quantum correlations. These measures are  
based on different metrics quantifying the minimum distance of the quantum state 
from the set of all possible classical-quantum states
\cite{dakic_geom,qd_geom,qd_geom_adesso,qd_geom_one,qd_geom_prob,qd_geom_use,bell_x_onenorm}. 
Although a collection of distance metrics have been
used to characterize geometric measures of quantum correlations, it has been shown that 
out of all the Schatten $p$-norm distances, 
only the one-norm distance has properties that are rather similar to the QD as well as QWD 
\cite{qd_geom_prob,qd_geom_one,bell_x_onenorm}.  But due to the difficulty in optimizing the measure, 
analytically closed forms of one-norm geometric discord has been obtained only for some special 
types of states, e.g., Bell diagonal states, and the X states \cite{bell_x_onenorm}.
% is fit to appropriately define geometric 
% measures 
% Note here that although closed form analytical expression of some of the distance based measures
% are available in the case of some very specific class of states \cite{qd_geom_adesso}, 
% Similar to The definitions of all such measures 
% are based on optimization over a complete set of quantum states having certain 
% properties. This makes analytical determination 
% of such measures in the case of arbitrary bipartite mixed quantum states almost intractable. 
Our methodology, along with the traditional QD and QWD, is applicable also to the geometric measures that 
require an optimization. Motivated by the usefulness of QD and QWD in certain quantum protocols 
\cite{nl_w_ent,opt_det,dqci,op_int}, we choose these measures for the purpose of demonstration.

\section{Two-qubit systems}
\label{twoqubit}

In the case of a $\mathbb{C}^{2}_{A}\otimes\mathbb{C}_{B}^{2}$ system where each of the subsystems consists of a single qubit only,  
the rank-$1$ projection measurements are of the form $\{\Pi^{A}_{k}=U|k\rangle\langle k|U^{\dagger}$, 
$|k\rangle=|0\rangle,|1\rangle\}$, where $U$, 
a local unitary operator in $SU(2)$, can be parametrized using two real parameters, $\theta$, and $\phi$, as  
\begin{eqnarray}
 U=\left( 
\begin{array}{cc}
\cos\frac{\theta}{2} & \sin\frac{\theta}{2} e^{i\phi}\\ 
-\sin\frac{\theta}{2} e^{-i\phi} & \cos\frac{\theta}{2} 
\end{array}
\right).
\label{pi12}
\end{eqnarray} 
Here, $0 \leq \theta \leq \pi$, $0 \leq \phi < 2\pi$, and $\{|0\rangle,|1\rangle\}$ denotes the computational basis in $\mathbb{C}^{2}$.
Note that $\theta$ and $\phi$ can be identified as the azimuthal and the polar angles, respectively, in the 
Bloch sphere representation of a qubit. 
% The projection basis of the rank-$1$ projectors are spanned over the surface of the 
% Bloch sphere. 
Let us define a parameter transformation, $f_{\theta}=\cos\theta$, so that $-1\leq f_{\theta}\leq 1$. 
Here and throughout this paper, whenever we need to perform an optimization over all rank-$1$ projection measurements on a qubit to
evaluate a given quantum correlation, $Q$, we choose the parameters $f_{\theta}$ and $\phi$ uniformly in $[-1,1]$ and $[0,2\pi]$
respectively. 
% we perform the optimization
% of quantum correlation measure, $Q$, for a two-qubit state, $\rho_{AB}$, over the $\{f_{\theta},\phi\}$ parameter space.
An earmarked set, in the present case, is equivalent to a subset of the complete set of allowed values of $f_{\theta}$ and $\phi$.

\subsection{Mixed states with different ranks}
\label{gen_mix}

First, we discuss the case of general two-qubit mixed states of different ranks.
As described above, an optimal set of $(f_{\theta},\phi)$ values define the optimal projection measurement 
for the computation of the fixed quantum correlation measure, $Q$.  
Let us consider the probability, $p_{r}$, that the optimal values of the real parameters, $f_{\theta}$ and $\phi$, 
for a fixed measure of quantum correlation, $Q$, of a randomly chosen two-qubit mixed state of rank $r$, lie in 
$(f_{\theta}, f_{\theta}+df_{\theta})$, and $(\phi,\phi+d\phi)$, respectively. 
The fact that the real parameters, $f_{\theta}$ and $\phi$, are independent of each other suggests that 
\begin{eqnarray}
p_{r}=P_{r}(f_{\theta},\phi)df_{\theta} d\phi=P_{r}^1(f_{\theta})df_{\theta} P_{r}^2(\phi)d\phi,
\label{para_pdf}
\end{eqnarray} 
which allows one to investigate the two probability density functions (PDFs), $P_{r}^1(f_{\theta})$, and $P_{r}^2(\phi)$, 
independently. Here, $P_{r}^1(f_{\theta})df_{\theta}$ denotes the probability that irrespective of the optimal value of $\phi$, 
the optimal value of $f_{\theta}$ lies between $f_{\theta}$ and $f_{\theta}+df_{\theta}$ for the fixed quantum correlation measure, 
$Q$, calculated for a two-qubit mixed quantum state of rank $r$. A similar definition holds for the probability 
$P_{r}^2(\phi)d\phi$ also. 

In the case of two-qubit systems, an uniform distribution of the projection measurements in the measurement space corresponds
to the uniform distribution of the $(f_\theta,\phi)$ points on the surface of the Bloch sphere. It is, therefore, reasonable to expect
that the PDFs, $P_{r}^1(f_{\theta})df_{\theta}$ and $P_{r}^2(\phi)d\phi$, correspond to uniform distributions over the allowed ranges
of values of $f_\theta$ and $\phi$. To verify this numerically, 
we consider QD and QWD as the chosen measures of quantum correlation.
The corresponding $P_{r}^1(f_{\theta})$, and $P_{r}^2(\phi)$ in the case of two-qubit mixed states having NPPT or having 
PPT for both QD and QWD are determined by 
generating $5\times10^5$ states Haar uniformly for each value of $r=2$, $3$, and $4$. We find that in the case of QD as well as QWD, 
both $P_{r}^1(f_{\theta})$, and $P_{r}^2(\phi)$ are uniform distributions over the entire ranges of corresponding parameters, 
$f_{\theta}$ and $\phi$, irrespective of the rank of the state as well as whether the state is NPPT or PPT.  
Note here that for two-qubit mixed states of rank-$2$, almost all states are NPPT while the PPT states form a set of measure zero 
\cite{r2ppt}, which can also be verified numerically. However, 
in the case of $r=3$ and $4$, non-zero volumes of PPT states are found. In $\mathbb{C}^{2}\otimes\mathbb{C}^{2}$ systems, 
all NPPT states are entangled while PPT states form the set of separable states \cite{volume}.

The fact that all the rank-$1$ projection measurements are equally probable makes the qualitative features of $Q_c$ depend only on the 
geometrical structure of the earmarked set, $\mathcal{S}_E$, and not on the actual location of the elements of the set on the Bloch sphere. 
% makes the choice of the earmarked set arbitrary. It 
% depends completely on the constraint over the complete set of projectors, which, in turn, depends on the situation in hand. 
% There may be a number of plausible ways in which one can choose an earmarked set, $S_{E}$, for a fixed quantum correlation measure, 
% $Q$. 
In the following, we consider four distinct choices of the set $\mathcal{S}_{E}$ for both QD and QWD in the case of two-qubit 
mixed states with ranks $r=2,3,4$, and discuss the corresponding scaling of the average VE.

\subsubsection*{\textbf{Case 1: \texorpdfstring{$\mathcal{S}_{E}$}{} with \texorpdfstring{($f_\theta,\phi$)}{} 
distributed over a circle on the Bloch sphere}}

We start by constructing the earmarked set with projection measurements such that the corresponding $(f_\theta,\phi)$  lies on the circle 
of intersection of a fixed plane with the Bloch sphere.
% basis on the perimeter of a fixed plane rather 
% than considering them spanned over the entire surface of the Bloch sphere. 
Let us assume that the corresponding VE resulting from the restricted optimization of 
$Q$, for an arbitrary two-qubit mixed state  
of rank $r$ is given by $\varepsilon_{n}^{r}$, $n$ being the size of $\mathcal{S}_{E}$, and where the corresponding $n$ points on the 
circle are symmetrically placed. 
Let us also assume that the CQC calculated by constrained optimization over the set $\mathcal{S}_{E}$ with $n\rightarrow\infty$ 
is given by $Q_{c}^\prime$ and the corresponding VE, called the ``asymptotic error'', 
is $\varepsilon_{\infty}^{r}=|Q_{c}^\prime-Q_{a}|$. 
To investigate how fast $\varepsilon_{n}^{r}$ reaches $\varepsilon_{\infty}^{r}$ on average with 
increasing $n$, one must look into the variation of 
$\overline{\varepsilon}_{n}^{r}-\overline{\varepsilon}_{\infty}^{r}$ against $n$ for different values of $r$. 
Here, $\overline{\varepsilon}_{n}^{r}$ is the average value of the VE, $\varepsilon_{n}^{r}$, and is given by 
\begin{eqnarray}
 \overline{\varepsilon}_{n}^{r}=\int_{0}^{1}\varepsilon_{n}^{r}P_{n}^r(\varepsilon_{n}^{r})d\varepsilon_{n}^{r},
 \label{ave}
\end{eqnarray}
with $P_{n}^r(\varepsilon_{n}^{r})d\varepsilon_{n}^{r}$ being the  probability that for an arbitrary two-qubit mixed state with 
rank $r$, the VE lies between $\varepsilon_{n}^{r}$ and $\varepsilon_{n}^{r}+d\varepsilon_{n}^{r}$ 
when $Q_{c}$ is calculated over $\mathcal{S}_{E}$ of size $n$ defined on the chosen plane. 
A similar definition holds for the average asymptotic VE,  $\overline{\varepsilon}_{\infty}^{r}$, and the PDF, 
$P^r_{\infty}(\varepsilon_{\infty}^{r})$,  in the limit $n\rightarrow\infty$, where 
\begin{eqnarray}
 \overline{\varepsilon}_{\infty}^{r}=\int_{0}^{1}\varepsilon_{\infty}^{r}P_{\infty}^r(\varepsilon_{\infty}^{r})d\varepsilon_{\infty}^{r}.
 \label{asym_ave}
\end{eqnarray}

\begin{figure}
  \includegraphics[scale=0.3]{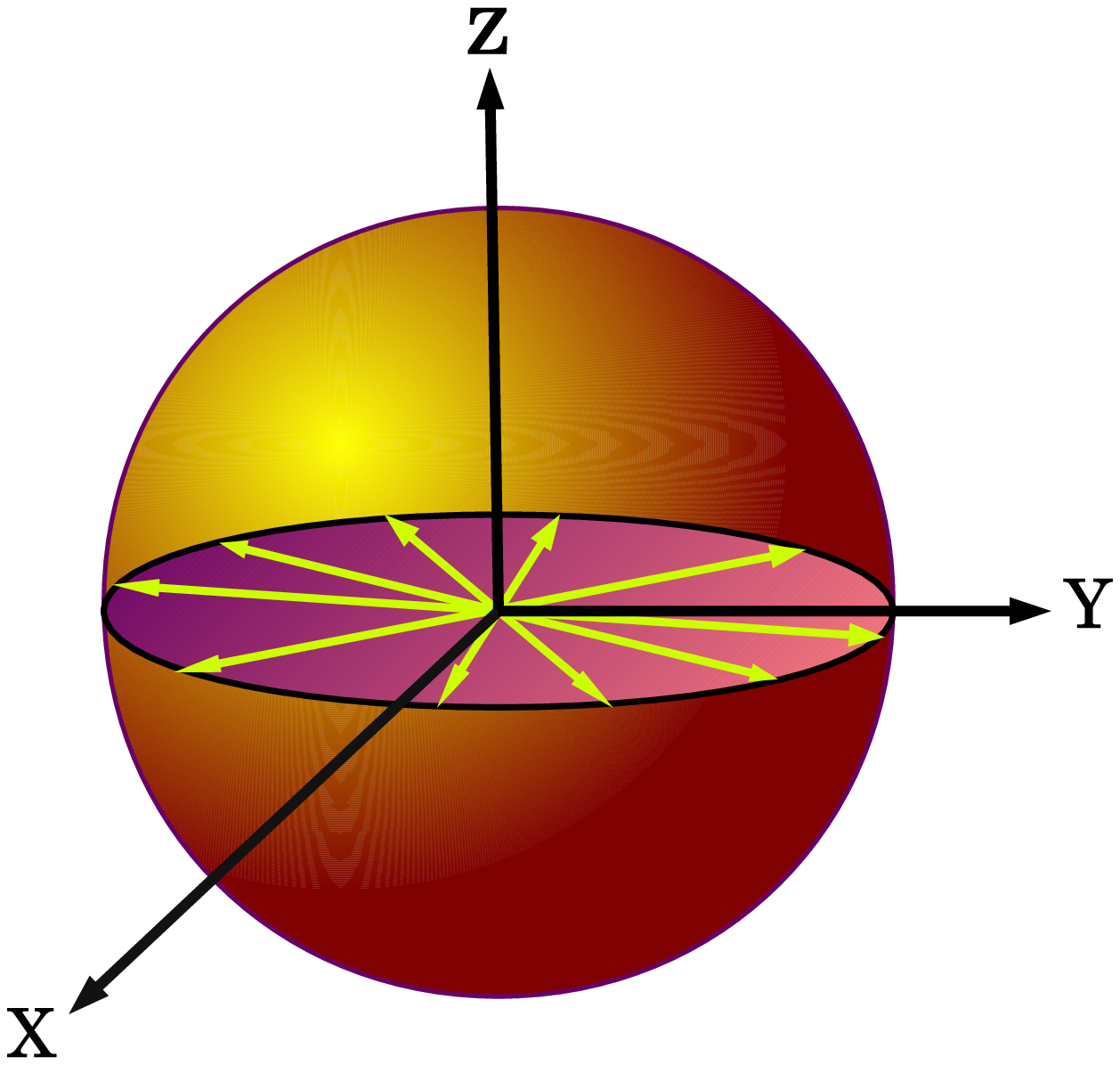} 
  \includegraphics[scale=0.3]{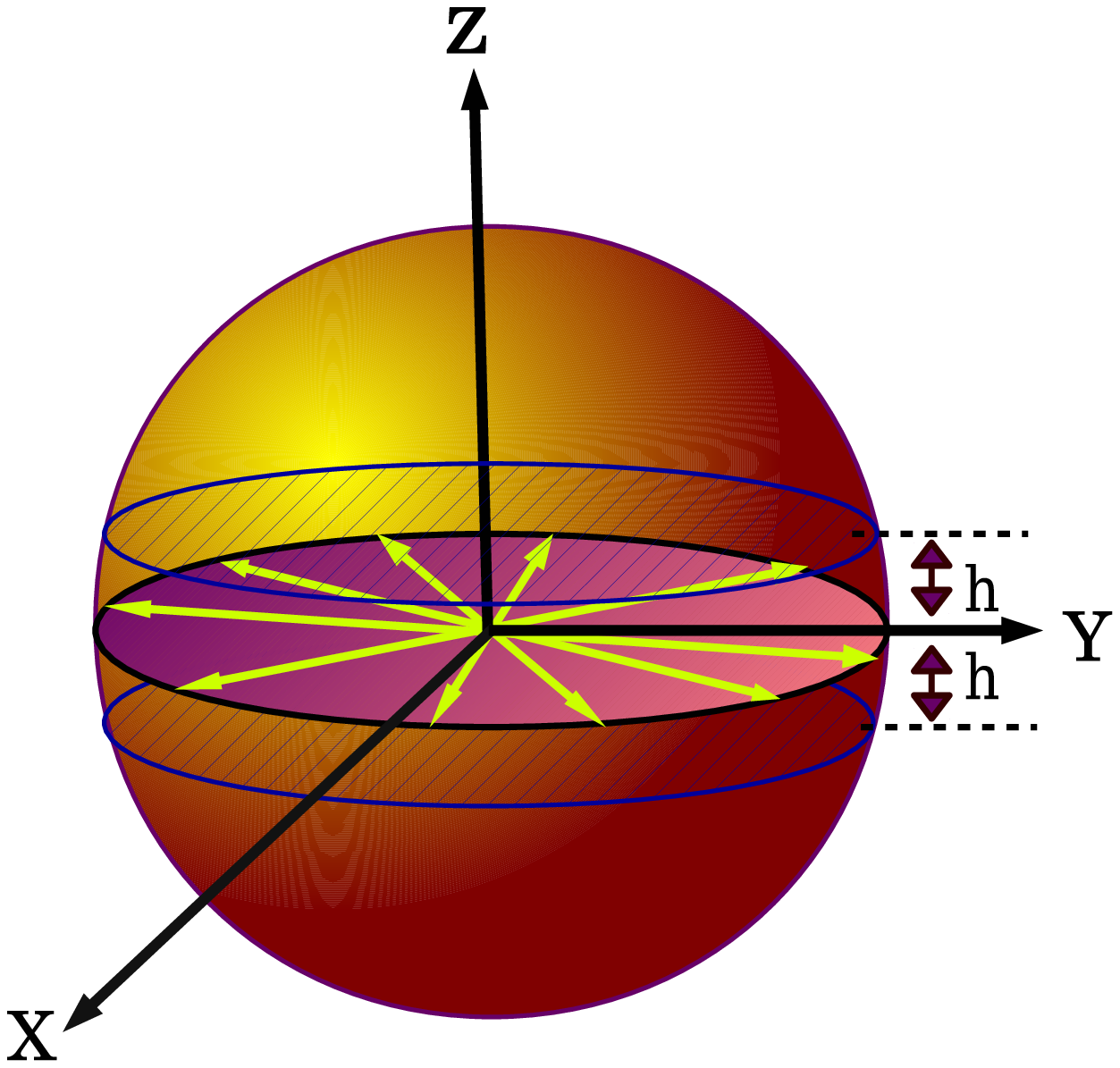}

 \textbf{(a)}\hspace{3.775cm}\textbf{(b)}
 \caption{(Color online.) (a) Schematic representation of the earmarked set confined on the circle defined by the intersection of the 
 $(x,y)$ plane, fixed by $f_\theta=0$, and the Bloch sphere. The plane is specified by the eigenbases of $\sigma^x$ and $\sigma^y$. 
 (b) Schematic representation of the earmarked set confined on a set of circles, defined by the intersections of a set of planes and the 
 Bloch sphere. The planes are considered to be symmetrically placed on either side of the $(x,y)$ plane fixed by $f_{\theta}=0$.}
 \label{schematic}
\end{figure}

% \begin{figure}[t!]
% %     \centering
%     \begin{subfigure}[t]{0.5\linewidth}
% %         \centering
%         \includegraphics[scale=0.3]{Figures/Fig1.eps}
%         \caption{}
%     \end{subfigure}%
%     \begin{subfigure}[t]{0.5\linewidth}
% %         \centering
%         \includegraphics[scale=0.3]{Figures/Fig2.eps}
%         \caption{}
%     \end{subfigure}
%     \caption{(Color online.) (a) Schematic representation of the earmarked set confined on the circle defined by the intersection of the 
%  $(x,y)$ plane, fixed by $f_\theta=0$, and the Bloch sphere. The plane is specified by the eigenbases of $\sigma^x$ and $\sigma^y$. 
%  (b) Schematic representation of the earmarked set confined on a set of circles, defined by the intersections of a set of planes and the 
%  Bloch sphere. The planes are considered to be symmetrically placed on either side of the $(x,y)$ plane fixed by $f_{\theta}=0$.}
%  \label{schematic}
% \end{figure}

We consider two different ways in which the plane is chosen. 
\textbf{(a)} We fix a value of $f_{\theta}=f_{\theta}^{\prime}$ such that the corresponding states on the Bloch sphere 
are given by 
$|\xi\rangle=\cos\frac{\cos^{-1}f_{\theta}^{\prime}}{2}|0\rangle+e^{i\phi}\sin\frac{\cos^{-1}f_{\theta}^{\prime}}{2}|1\rangle$, 
where $\phi$ acts as the spanning parameter. \textbf{(b)} In the second option, we fix the value of 
$\phi=\phi^{\prime}$ while vary $f_\theta$ with the corresponding states $|\xi\rangle$. 
We consider both the scenarios, and investigate the scaling of the corresponding average VEs for both QD and QWD for arbitrary
two-qubit mixed states of different ranks.

\begin{figure}
 \includegraphics[scale=0.675]{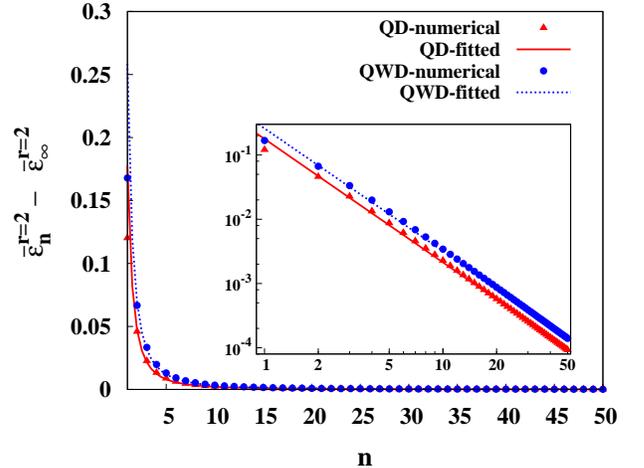}
 \caption{(Color online.) Variation of $\overline{\varepsilon}_{n}^{r=2}-\overline{\varepsilon}_{\infty}^{r=2}$ as a function of 
 $n$ in the case of QD and QWD. (Inset) Linear variation of $\overline{\varepsilon}_{n}^{r=2}-\overline{\varepsilon}_{\infty}^{r=2}$ 
 as a function of $n$ in the log-log graph for QD and QWD, where the variation is given by Eq. (\ref{powerlaw}).
 The numerical data is represented by points while the fitted curve is given by solid lines.
 The fitting parameters are estimated as $\kappa=1.77\times10^{-1}\pm1.22\times10^{-2}$, and 
$\tau=-1.92\pm1.11\times10^{-2}$ with $\varepsilon_{\infty}^{r=2}=1.21\times10^{-1}$ in the case of QD, whereas for QWD, 
$\kappa=2.58\times10^{-1}\pm1.78\times10^{-2}$, $\tau=-1.91\pm1.23\times10^{-2}$, and 
$\varepsilon_{\infty}^{r=2}=1.66\times10^{-1}$. In the main figure, the abscissa is dimensionless, while the 
quantities $\overline{\varepsilon}_n^{r=2}$ and $\overline{\varepsilon}_\infty^{r=2}$ are in bits. In the inset, the ordinate is in the 
natural logarithm of $\overline{\varepsilon}_n^{r=2}-\overline{\varepsilon}_\infty^{r=2}$, and the $x$ axis is in the natural 
logarithm of $n$.}
 \label{rank2}
\end{figure}

\noindent\textbf{(a) Fixed value of $f_{\theta}$:} Unless otherwise stated, here and throughout this paper, we shall fix a plane 
by assigning a value to $f_\theta$. 
For the purpose of demonstration, we choose $f_{\theta}=0$, fixing the $(x,y)$ plane defined 
by the eigenbasis of the Pauli matrices $\sigma^{x}$ and $\sigma^{y}$, where an arbitrary projection basis can be written as 
$|\xi\rangle=\frac{1}{\sqrt{2}}(|0\rangle+e^{i\phi}|1\rangle)$. The earmarked set of size $n$ on the perimeter of the circle 
of intersection of the $(x,y)$ plane and the Bloch sphere can 
be generated by a set of $n$ projectors of the form $U|k\rangle\langle k|U^{\dagger}$, $|k\rangle=|0\rangle,|1\rangle$, 
obtained by using a
fixed $f_{\theta}=0$, and $n$ equispaced divisions of the entire range of $\phi$. Here, the form of $U$ is given in Eq. (\ref{pi12}).
The situation is depicted in Fig. \ref{schematic}(a). To determine the PDFs, $P_{n}^r(\varepsilon_{n}^r)$ and 
$P_{\infty}^r(\varepsilon_{\infty}^r)$, we Haar uniformly generate 
$5\times10^5$ random two-qubit mixed states of rank $r=2,3$, and $4$ each. The corresponding average VE, 
$\overline{\varepsilon}_{n}^{r}$, and average asymptotic VE, $\overline{\varepsilon}_{\infty}^{r}$,
in the case of both QD and QWD are determined using Eqs. (\ref{ave}) and (\ref{asym_ave}) respectively. 
For both the quantum correlation measures, the quantity 
$\overline{\varepsilon}_{n}^{r}-\overline{\varepsilon}_{\infty}^{r}$ is found to have a 
power-law decay with the size, $n$, of the earmarked set, on the log-log scale for all values of $r$. 
One can determine the functional dependence of $\overline{\varepsilon}_{n}^{r}$ over $n$ as 
\begin{eqnarray}
\overline{\varepsilon}_{n}^{r}=\overline{\varepsilon}_{\infty}^{r}+\kappa n^{-\tau},
\label{powerlaw}
\end{eqnarray}
where the fitting constant, $\kappa$, and the scaling exponent, $\tau$, are estimated from the numerical data.   Fig. \ref{rank2}  
shows the variations of $\overline{\varepsilon}_{n}^{r}-\overline{\varepsilon}_{\infty}^{r}$ with $n$
for both QD and QWD in the case of rank-$2$ two-qubit states. 
The insets in Fig. \ref{rank2} shows the corresponding variations of 
$\overline{\varepsilon}_{n}^{r}-\overline{\varepsilon}_{\infty}^{r}$ with increasing $n$ in the log-log scale.

\begin{table}
 \begin{tabular}{|c|}
 \hline
  \cellcolor{blue!25} \textbf{QD} \\
 \hline 
 \begin{tabular}{c|c|c}
  \textcolor{blue}{$r$} & \textcolor{red}{NPPT} & \textcolor{red}{PPT} \\
  \hline
  
  \footnotesize3\normalsize & \shortstack{\footnotesize$\kappa=1.45\times10^{-1}\pm9.97\times10^{-3}$\normalsize\\
  \footnotesize$\tau=1.93\pm1.05\times10^{-2}$\normalsize\\
  \footnotesize$\overline{\varepsilon}^{r=3}_{\infty}=9.56\times10^{-2}$\normalsize}  
  &\shortstack{\footnotesize$\kappa=1.04\times10^{-1}\pm7.10\times10^{-2}$\normalsize\\ 
  \footnotesize$\tau=1.94\pm1.01\times10^{-2}$\normalsize\\
  \footnotesize$\overline{\varepsilon}^{r=3}_{\infty}=6.98\times10^{-2}$\normalsize} \\
  \hline
  \footnotesize4\normalsize & \shortstack{\footnotesize$\kappa=1.21\times10^{-1}\pm8.32\times10^{-3}$\normalsize\\ 
  \footnotesize$\tau=1.94\pm1.02\times10^{-2}$\normalsize\\
  \footnotesize$\overline{\varepsilon}^{r=4}_{\infty}=7.80\times10^{-2}$\normalsize} 
  &\shortstack{\footnotesize$\kappa=8.67\times10^{-2}\pm5.96\times10^{-3}$\normalsize\\
  \footnotesize$\tau=1.94\pm9.90\times10^{-3}$\normalsize\\
  \footnotesize$\overline{\varepsilon}^{r=4}_{\infty}=5.78\times10^{-2}$\normalsize}\\
 \end{tabular}\\
 \hline
 \cellcolor{green!25} \textbf{QWD} \\
 \hline
  \begin{tabular}{c|c|c}
  \textcolor{blue}{$r$} & \textcolor{red}{NPPT} & \textcolor{red}{PPT} \\
  \hline  
  \footnotesize3\normalsize & \shortstack{\footnotesize$\kappa=1.89\times10^{-1}\pm1.30\times10^{-2}$\normalsize\\
  \footnotesize$\tau=1.93\pm1.13\times10^{-2}$\normalsize\\
  \footnotesize$\overline{\varepsilon}^{r=3}_{\infty}=1.18\times10^{-1}$\normalsize}  
  &\shortstack{\footnotesize$\kappa=1.52\times10^{-1}\pm1.04\times10^{-2}$\normalsize\\
  \footnotesize$\tau=1.93\pm1.08\times10^{-2}$\normalsize\\
  \footnotesize$\overline{\varepsilon}^{r=3}_{\infty}=9.64\times10^{-2}$\normalsize} \\  
  \hline
  \footnotesize4\normalsize & \shortstack{\footnotesize$\kappa=1.50\times10^{-1}\pm1.03\times10^{-2}$\normalsize\\ 
  \footnotesize$\tau=1.93\pm1.07\times10^{-2}$\normalsize\\ 
  \footnotesize$\overline{\varepsilon}^{r=4}_{\infty}=9.23\times10^{-2}$\normalsize} 
  &\shortstack{\footnotesize$\kappa=1.20\times10^{-1}\pm8.26\times10^{-3}$\normalsize\\ 
  \footnotesize$\tau=1.94\pm1.04\times10^{-2}$\normalsize\\
  \footnotesize$\overline{\varepsilon}^{r=4}_{\infty}=7.52\times10^{-2}$\normalsize}\\
 \end{tabular}\\ 
 \hline 
 \end{tabular}
\caption{Values of the fitting constant, $\kappa$, scaling exponent, $\tau$, and asymptotic error, 
$\overline{\varepsilon}_{\infty}^{r}$ in the case of NPPT as well as PPT two-qubit mixed states of rank $r=3$ and $4$. 
Here, QD and QWD are considered as quantum correlation measures, and the earmarked set is fixed on the circle defined by the intersection 
of a plane fixed by $f_{\theta}=0$, and the Bloch sphere. The corresponding power-law variations of 
$\overline{\varepsilon}_{n}^{r=2}-\overline{\varepsilon}_{\infty}^{r=2}$ with $n$ are depicted in Fig. \ref{rank34}.}
\label{tab}
\end{table}

\begin{figure*}
 \includegraphics[width=0.7\textwidth]{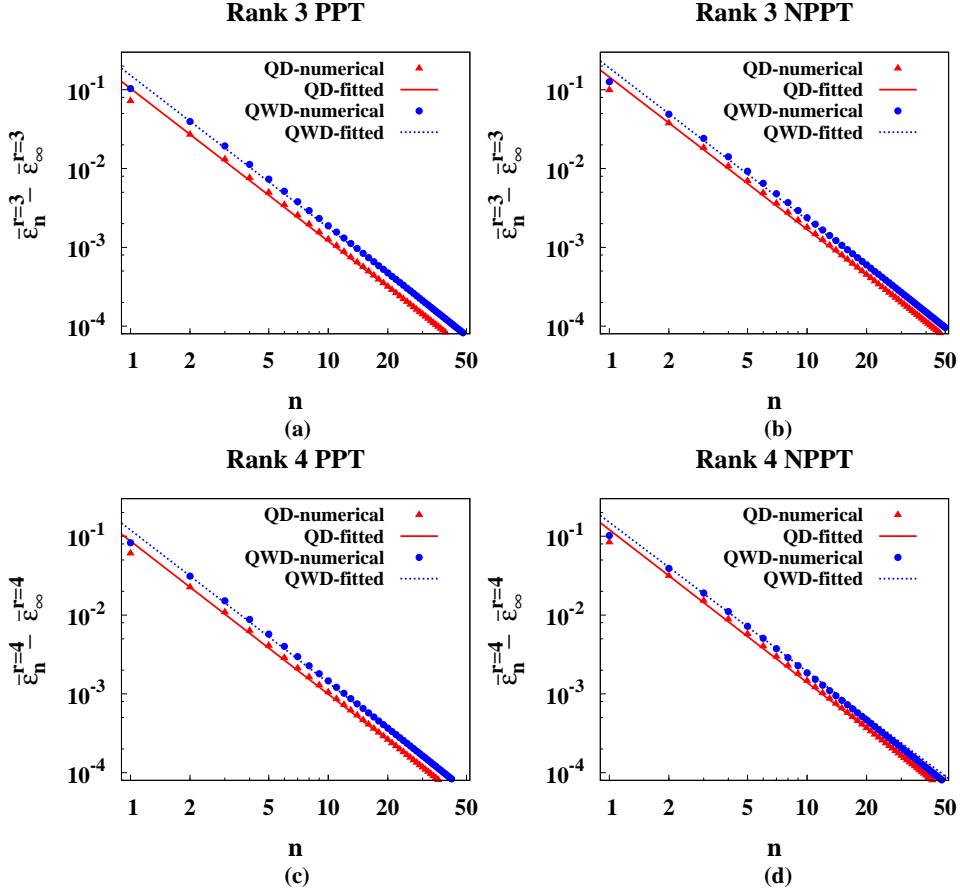}
 \caption{(Color online.) Variations of $\overline{\varepsilon}_{n}^{r}-\overline{\varepsilon}_{\infty}^{r}$ 
 as a function of $n$, in log-log scale, for $r=3$ and $4$ in the case of NPPT and PPT two-qubit mixed states using QD and QWD.
 The numerical data is represented by points while the fitted curve is given by solid lines. The corresponding 
 values of the fitting constant, $k$, scaling exponent, $\tau$, and asymptotic error, $\overline{\varepsilon}_{\infty}^{r}$,
 are tabulated in Table \ref{tab}. The ordinates of all the figures are in natural logarithm of 
 $\overline{\varepsilon}_{n}^{r}-\overline{\varepsilon}_{\infty}^{r}$, with $\overline{\varepsilon}_n^r$ and $\overline{\varepsilon}_\infty^r$ 
 individually being in bits, while the abscissa is in natural logarithm of the cardinality of the earmarked set.}
 \label{rank34}
\end{figure*}

The variations of $\overline{\varepsilon}_{n}^{r}-\overline{\varepsilon}_{\infty}^{r}$ with $n$ in the log-log scale using 
both QD and QWD with NPPT as well as PPT two-qubit mixed states of rank $r=3$ and $r=4$ are shown separately in Fig. \ref{rank34}. 
The corresponding exponents and fitting parameters are quoted in Table \ref{tab}. 
Note that although the exponent, $\tau$, in the case of QD and QWD has 
equal values up to the first decimal point for all the cases (for NPPT and PPT states with different ranks), 
the average asymptotic VE, $\overline{\varepsilon}_{\infty}^{r}$, is larger in the case of QWD in comparison to that for QD. 
This is reflected  in the fact that the graph for QWD is above the graph for QD, as shown in Fig. \ref{rank2} and \ref{rank34}.
Note also that $\overline{\varepsilon}_{n}^{r}$ is less in the case of PPT states than the NPPT states, when two-qubit
states of a fixed rank are considered.

Note that in the above example, we have fixed the plane by fixing $f_{\theta}=0$, which corresponds to a great circle on the 
Bloch sphere. One can also fix a great circle on the Bloch sphere by fixing any value of $\phi$ in its allowed range of values. 
Earmarked sets defined over any such great circle on the Bloch sphere have similar scaling properties of the average VE as long as 
the points corresponding to the projection measurements belonging to the set $\mathcal{S}_E$ are distributed uniformly over
the circle. However, for different distribution, one can obtain different scaling exponents and fitting parameters. We 
shall shortly
discuss one such example. 

For $f_{\theta}\neq0$, smaller circles over the Bloch spheres are obtained. One can 
also investigate the scaling behaviour of the average VE in the case of earmarked sets having elements corresponding to points 
distributed over these smaller circles by using $\phi$ as the spanning parameter. However, 
different values for the scaling exponents and fitting 
parameters are obtained as $|f_{\theta}|\rightarrow1$.

% The results presented here are obtained by restricting the earmarked set on the perimeter of the $(x,y)$ plane. However, 
% the qualitative results as well as the exponents remain unchanged when the plane is fixed by any arbitrary value of $f_{\theta}$.  

\noindent\textbf{(b) Fixed value of $\phi$:}
Next, we consider the scenario where the distribution of the $(f_\theta,\phi)$ points corresponding to the projection measurements
constituting the set $\mathcal{S}_{E}$ on the circle of our choice is different (i.e., non-uniform) than the previous example. This may 
happen due to restrictions imposed by apparatus during experiment, or other relevant physical constraints. 
As before, we choose the plane by fixing $\phi=0$, 
thereby confining the earmarked set on the great circle representing the intersection of the $(x,z)$ plane defined by the 
eigenbases of $\sigma^{x}$ and $\sigma^{z}$, and the Bloch sphere. 
However, to demonstrate the effect of such non-uniformity over the scaling parameters, we consider 
the $(f_\theta,\phi)$ points corresponding to the $n$ elements in $\mathcal{S}_{E}$ by $n$ equal divisions of the entire 
range of $f_{\theta}$. Note that the current choice of $f_\theta$ as the spanning parameter leads to a different (non-uniform)
distribution of the points corresponding to the projection measurements in $\mathcal{S}_E$ on the chosen circle.

Similar to the previous case of $f_\theta=0$, one can also study the scaling of average VE 
by defining $\overline{\varepsilon}_{n}^{r}$ and $\overline{\varepsilon}_{\infty}^{r}$ corresponding to the present case.
% considering 
% QD and QWD in the case of both NPPT as well as PPT two-qubit mixed states of different ranks, 
The variation of $\overline{\varepsilon}_{n}^{r}-\overline{\varepsilon}_{\infty}^{r}$ with $n$, as in the previous case, is given by 
Eq. (\ref{powerlaw}), only with different values of fitting constant, scaling exponent, and average asymptotic error. 
In the case of two-qubit mixed states of rank-$2$, 
the appropriate parameter values for QD are found to be $\kappa=1.30\times10^{-1}\pm1.08\times10^{-3}$, $\tau=1.47\pm2.7\times10^{-3}$,
with $\overline{\varepsilon}_{\infty}^{r=2}=1.21\times10^{-1}$. In the case of states with $r=3$ and $4$, the 
values of $\kappa$, $\tau$, and $\overline{\varepsilon}_{\infty}^{r}$ in the case of QD are tabulated in Table \ref{tab2}.
% when QD is considered as the correlation measure. 
Note that the scaling exponents are different from those found in case of a fixed $f_{\theta}$. 
However, the fact that the average asymptotic VE is less in the case of PPT states compared to that of NPPT states 
remains unchanged even in the present scenario.

\begin{table}
 \begin{tabular}{|c|}
 \hline
  \cellcolor{blue!25} \textbf{QD} \\
 \hline 
 \begin{tabular}{c|c|c}
  \textcolor{blue}{$r$} & \textcolor{red}{NPPT} & \textcolor{red}{PPT} \\
  \hline  
  \footnotesize3\normalsize & \shortstack{\footnotesize$\kappa=1.04\times10^{-1}\pm5.60\times10^{-4}$\normalsize\\ 
  \footnotesize$\tau=1.47\pm1.70\times10^{-3}$\normalsize\\ 
  \footnotesize$\overline{\varepsilon}^{r=3}_{\infty}=9.54\times10^{-2}$\normalsize}   
  & \shortstack{\footnotesize$\kappa=7.68\times10^{-2}\pm4.3\times10^{-4}$\normalsize\\ 
  \footnotesize$\tau=1.48\pm1.80\times10^{-3}$\normalsize\\
  \footnotesize$\overline{\varepsilon}^{r=3}_{\infty}=6.95\times10^{-2}$\normalsize} \\  
  \hline
  \footnotesize4\normalsize & \shortstack{\footnotesize$\kappa=8.95\times10^{-2}\pm9.6\times10^{-4}$\normalsize\\ 
  \footnotesize$\tau=1.48\pm3.4\times10^{-3}$\normalsize\\ 
  \footnotesize$\overline{\varepsilon}^{r=3}_{\infty}=7.81\times10^{-2}$\normalsize}   
  &\shortstack{\footnotesize$\kappa=6.23\times10^{-2}\pm2.8\times10^{-4}$\normalsize\\ 
  \footnotesize$\tau=1.48\pm1.4\times10^{-3}$\normalsize\\
  \footnotesize$\overline{\varepsilon}^{r=4}_{\infty}=5.79\times10^{-2}$\normalsize} \\
 \end{tabular}\\
 \hline 
 \end{tabular}
\caption{Values of the fitting constant, $\kappa$, scaling exponent, $\tau$, and asymptotic error, 
$\overline{\varepsilon}_{\infty}^{r}$ in the case of NPPT as well as PPT two-qubit mixed states of rank $r=3$ and $4$. 
Here, quantum correlation is quantified by QD, and the earmarked set is fixed on the circle of intersection of 
a plane fixed by $\phi=0$, and the Bloch sphere. The $n$ points corresponding to the projection measurements in $\mathcal{S}_E$ 
are distributed over the circle by taking into account $n$ equispaced division of the range $[-1,1]$ of the spanning parameter 
$f_\theta$.}
\label{tab2}
\end{table}

\begin{figure}
 \includegraphics[scale=0.6]{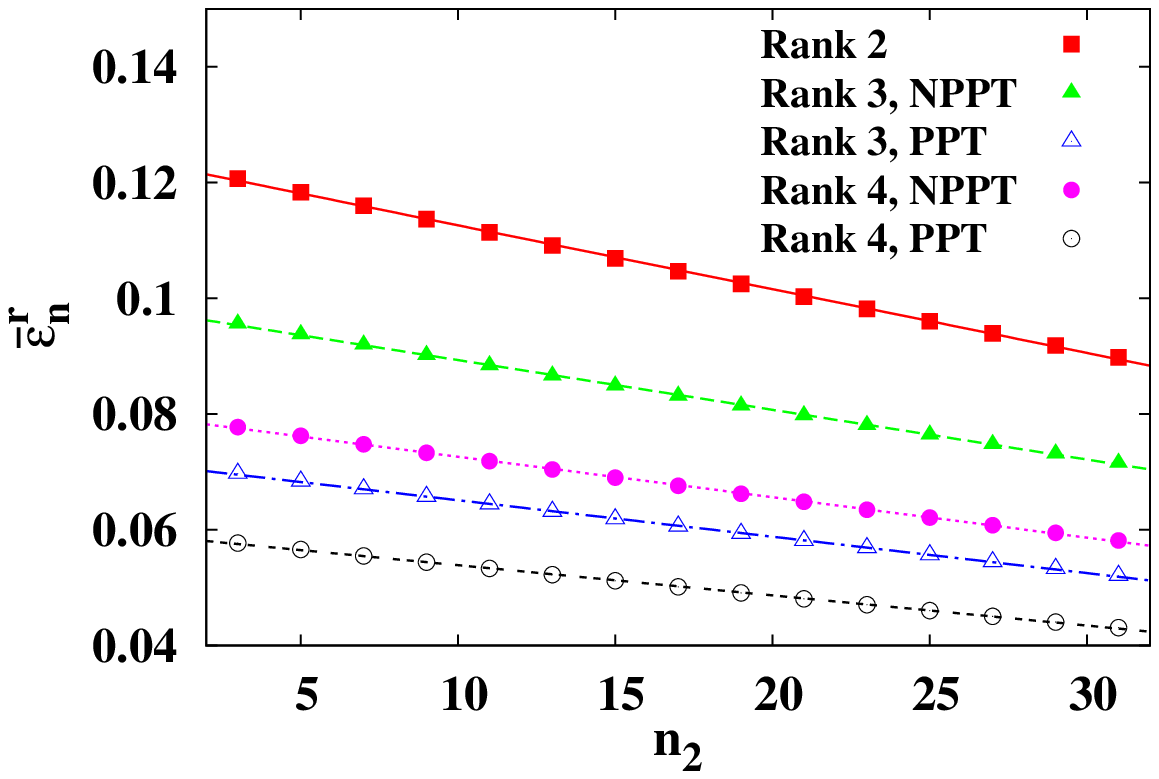}
 \caption{(Color online.) Linear variation of $\overline{\varepsilon}_{n}^{r}$ as a function of $n_{2}$ with $n_{1}=10$ ($n=n_{1}n_{2}$)
  for QD. 
  The earmarked set, in this case, is chosen on the perimeters of a set of discs in the Bloch sphere, starting from the 
  disc fixed by  $f_{\theta}=0$, and placing the additional discs symmetrically on either side of the $f_{\theta}=0$ disc.
  The numerical data obtained in each case is fitted to a straight line, $\overline{\varepsilon}_{n}^{r}=mn_{2}+c$. The fitting 
  parameter values are (i) $m=-1.10\times10^{-3},c=0.12$ (rank-$2$ states), (ii) $m=-8.59\times10^{-4},c=9.7\times10^{-2}$ 
  (rank-$3$ NPPT states), (iii) $m=-6.29\times10^{-4},c=7.14\times10^{-2}$ (rank-$3$ PPT states), 
  (iv) $m=-6.99\times10^{-4},c=7.96\times10^{-2}$ (rank-$4$ NPPT states), and 
  (v) $m=-5.22\times10^{-4},c=5.90\times10^{-2}$ (rank-$4$ PPT states). The abscissa is dimensionless, while the quantity 
  $\overline{\varepsilon}_n^r$ is in bits.} 
 \label{dis_lin}
\end{figure}

One can carry out similar investigation taking QWD as the chosen measure of quantum correlation. As in the previous case where 
$f_{\theta}=0$, here also the exponent in the case of QWD is found to be same with that for QD up to the first decimal place, 
although the average asymptotic error is higher.

% which can be explained by noting that in the current scenario, 
% the elements of the set $\mathcal{S}_{E}$ do not span the fixed plane in the same way as in the previous case. 

\subsubsection*{\textbf{Case 2: \texorpdfstring{$\mathcal{S}_{E}$}{} with \texorpdfstring{($f_\theta,\phi$)}{} on 
a collection of circles on the Bloch sphere}}

We now consider the situation where the projection measurements in the earmarked set are such that the corresponding $(f_\theta,\phi)$  
are not confined on the perimeter of a single fixed disc only, 
but are lying on the perimeters of a set of fixed discs (Fig. \ref{schematic}(b)). 
As discussed earlier, there are several ways in which a disc in the state
space can be fixed. For the purpose of demonstration, we achieve this by fixing the value of $f_{\theta}$. 
The size, $n$, of the set $\mathcal{S}_{E}$, depends on two quantities: \textbf{(i)} the number, $n_{1}$, of divisions of 
the allowed range of $\phi$ on any one of the discs, and 
\textbf{(ii)} the number, $n_{2}$, of discs that are considered for 
constructing the earmarked set. Evidently, $n=n_{1}n_{2}$.
Note that the value of $n_{1}$ is assumed to be constant for every disc, although 
one may consider, in principle, a varying number, $n_{1}^{i}$, such that $n=\sum_{i=1}^{n_{2}}n_{1}^{i}$.

\begin{table}[b]
 \begin{tabular}{|c|c|}
 \hline
 \cellcolor{blue!25} \textbf{QD} & \cellcolor{green!25} \textbf{QWD} \\
 \hline 
 \begin{tabular}{c|c|c}
  \textcolor{blue}{$r$} & \textcolor{red}{NPPT} & \textcolor{red}{PPT} \\
  \hline
  2 & \shortstack{$n_{1}=7$\\$n_{2}=9$} & \cellcolor{black!25} -- \\
  \hline
  3 & \shortstack{$n_{1}=6$\\$n_{2}=8$} & \shortstack{$n_{1}=5$\\$n_{2}=7$} \\
  \hline  
  4 & \shortstack{$n_{1}=5$\\$n_{2}=8$} & \shortstack{$n_{1}=5$\\$n_{2}=6$} \\
 \end{tabular}
 &
 \begin{tabular}{c|c|c}
  \textcolor{blue}{$r$} & \textcolor{red}{NPPT} & \textcolor{red}{PPT} \\
  \hline
  2 & \shortstack{$n_{1}=9$\\$n_{2}=11$} & \cellcolor{black!25} -- \\
  \hline
  3 & \shortstack{$n_{1}=7$\\$n_{2}=10$} & \shortstack{$n_{1}=6$\\$n_{2}=9$} \\
  \hline  
  4 & \shortstack{$n_{1}=6$\\$n_{2}=9$} & \shortstack{$n_{1}=5$\\$n_{2}=8$} \\
 \end{tabular} \\
 \hline 
 \end{tabular}
\caption{Values of $n_{1}$ and $n_{2}$ that are sufficient to obtain a value of average VE of the order of $10^{-3}$ in 
the case of NPPT and PPT 
two-qubit mixed states of different ranks, $r$, for QD and QWD. The values
correspond to the case where the earmarked set is confined on the surface of the Bloch sphere. The corresponding depiction 
is available in Fig. \ref{surf}.}
\label{tab3}
\end{table}

We demonstrate the situation with an example where an arbitrary disc is fixed by $f_{\theta}=f^\prime_\theta$, 
and a collection of additional discs positioned symmetrically with respect to the fixed disc is considered.
The fact that the number $n_{2}$ includes the fixed disc itself 
implies that $n_{2}$ is always an odd number.
In particular, the discs defining the set $\mathcal{S}_{E}$ can be marked with different values of 
$f_{\theta}$, given by $f_{\theta}^{j}=f_\theta^\prime\pm jh$, where $0\leq j\leq (n_{2}-1)/2$, and $h=2/(n_{2}-1)$. 
The set $\mathcal{S}_{E}$, in the present case, approaches the complete set of 
rank-$1$ projectors, $\mathcal{S}_{C}$, when both $n_{1}$ and $n_{2}$ tend to infinity. 
Similar to the previous case, one can study the variation of average VE with the increase in the size, $n$, of the set $\mathcal{S}_{E}$. 
However, unlike the previous case, in the situation where $n\rightarrow\infty$ is consequent to $(f_\theta,\phi)$ corresponding to the 
earmarked set being distributed over the entire Bloch sphere, 
% for $n_{1}\rightarrow\infty$ and $n_{2}\rightarrow\infty$
% together, 
the average asymptotic error, $\overline{\varepsilon}_{\infty}^{r}=0$ for all quantum correlation measures, since  
$Q_{c}\rightarrow Q_{a}$ in such cases.  

\begin{figure}
 \includegraphics[scale=0.45]{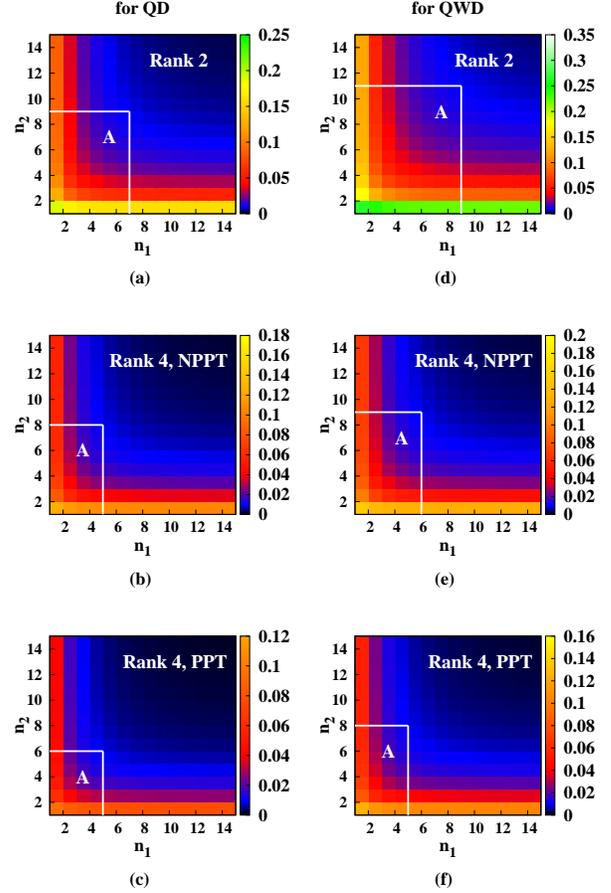}
 \caption{(Color online.) Variation of $\overline{\varepsilon}_{n}^{r}$ as a function of $n_{1}$ and $n_{2}$ ($n=n_{1}n_{2}$)
 in the case of NPPT and PPT two-qubit mixed states of rank $r=2$ and $r=4$, where QD and QWD are considered as quantum 
 correlation measures. The ranges of $n_{1}$ and $n_{2}$ marked by $A$ are sufficient to obtain a considerably low value of 
 $\overline{\varepsilon}_{n}^{r}$ ($\sim10^{-3}$). The area of the region decreases in the case of PPT states compared to NPPT states
 in the case of states with a fixed rank $r$. Also, the region is bigger in the case of QWD compared to that in QD, implying 
 a requirement of greater resource in the optimization of QWD. The corresponding ranges of $n_{1}$ and $n_{2}$ are tabulated in 
 Table \ref{tab3}. The different shades in the figure correspond to different values of $\overline{\varepsilon}_{n}^{r}$.
 All quantities plotted are dimensionless, except $\overline{\varepsilon}_n^r$, which is in bits.}
 \label{surf}
\end{figure}
 
Fixing $n_{1}$ to be a number for which the average VE calculated over an arbitrary disc in the set is considerably small,  
variation of the average VE, $\overline{\varepsilon}_{n}^{r}$, with increasing 
$n_{2}$, where $n=n_{1}n_{2}$, can be studied for QD and QWD. Here, the average is computed in a 
similar fashion as given in Eq. (\ref{ave}). However, $P_n^r(\varepsilon_{n}^r)$, in the present case, is obtained
by performing optimization of the corresponding quantum correlation measure using the current definition of $\mathcal{S}_{E}$.  
We fix $f^\prime_\theta=0$, and observe that in the case of QD, the average 
VE decreases linearly with increasing 
$n_{2}$ for all values of $n_{1}$ (See Fig. \ref{dis_lin}). 
This implies that for a fixed $n_{1}$, the spanning of the surface of the Bloch sphere by the 
perimeters of the discs in the set 
starting from $f_{\theta}=0$ occurs linearly with an increase in the value of $n_{2}$. Fig. \ref{dis_lin} depicts the variation of the 
average VE, 
$\overline{\varepsilon}_{n}^{r}$, as a function of $n_{2}$ for $n_{1}=10$, where the linear nature of the variation 
is clearly shown. 
% This result remains unchanged qualitatively if one chooses to start from a planed marked with a different 
% value of $f_{\theta}$. 
Similar results hold in the case of QWD also.

\begin{figure}
 \includegraphics[scale=0.45]{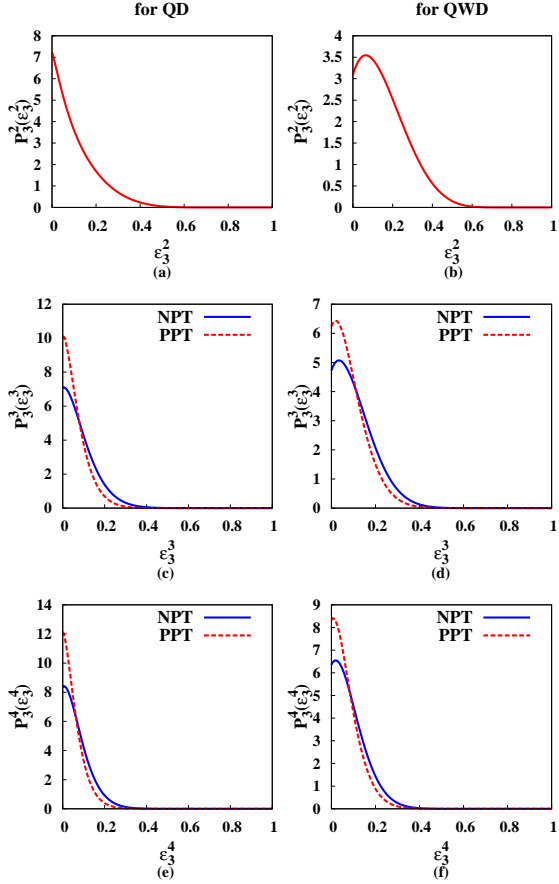}
 \caption{(Color online.) Profiles of the probability density function, $P_{3}^r(\varepsilon_{3}^r)$ against $\varepsilon_{3}^r$ 
 for NPPT and PPT two-qubit mixed states of different ranks, $r$. The distributions are sharply peaked around low values of 
 $\varepsilon_{3}^r$ in all cases considered. The earmarked set, in this case, is taken to be the triad. The ordinates in all the figures are 
 dimensionless, while the quantities $\overline{\varepsilon}_3^r$, $r=2,3,$ and $4$, are in bits.}
 \label{22err}
\end{figure}

% Note that in an alternative approach, one may start from a plane fixed by the value of $\phi$, say, 
% at $\phi^\prime$, and consider a fixed value of $n_{2}$, which is same for all the planes in the set. 
% The volume of the state space may be spanned by taking into account 
% $n_{1}-1$ additional planes symmetrically placed with respect to the plane marked with $\phi=\phi^\prime$, which are given by 
% $\phi^\prime\pm ih$, where $h=2\pi/(n_{1}-1)$. However, the all qualitative results remain unchanged with the latter choice. 

\subsubsection*{\textbf{Case 3: \texorpdfstring{$\mathcal{S}_{E}$}{} with \texorpdfstring{($f_\theta,\phi$)}{} 
on the surface of Bloch sphere}}

Let us now consider the situation where the bases in $\mathcal{S}_E$ are such that the corresponding $(f_\theta,\phi)$ are uniformly 
scattered over the entire surface of the Bloch sphere so that 
only $n_{1}$ and $n_{2}$ equal divisions of the entire range of $f_{\theta}$ and $\phi$, respectively, are 
allowed, leading to an earmarked set of size $n=n_{1}n_{2}$. Similar to the previous case, $\mathcal{S}_{E}\rightarrow\mathcal{S}_{C}$, 
and $\overline{\varepsilon}_{\infty}^{r}\rightarrow0$ when both $n_{1}$ and $n_{2}$ approach infinity. The average error, 
$\overline{\varepsilon}_{n}^{r}$, in the present case, is determined in a similar way as in Eq. (\ref{ave}) with the current 
description of $\mathcal{S}_{E}$. From the variation of $\overline{\varepsilon}_{n}^{r}$ as a function of $n_{1}$ 
and $n_{2}$ in the case of QD and QWD for two-qubit mixed NPPT and PPT states of rank $r=2,3$ and $4$,  
% The maximum available values of QD and QWD, with the increase in the rank of two-qubit mixed states, decrease. 
we estimate the minimum number of divisions, $n_{1}$, and $n_{2}$, required in order to converge on a sufficiently low value of 
$\overline{\varepsilon}_{n}^{r}$ ($\sim10^{-3}$) in each case. The corresponding values of $n_{1}$ and $n_{2}$ are 
given in Table \ref{tab3}. It is observed that 
with an increase in the rank of the state, the required size of the earmarked set, $n$, in order to converge on a sufficiently low 
value of average VE,  
\emph{reduces} for both QD and QWD. This feature is clearly depicted in Fig. \ref{surf} with a reduction in the area of the
region marked as $A$. Note also that the required size is smaller in the case of PPT states when compared to the NPPT states of same rank 
for a fixed quantum correlation measure.

\begin{figure*}
 \includegraphics[width=0.8\textwidth]{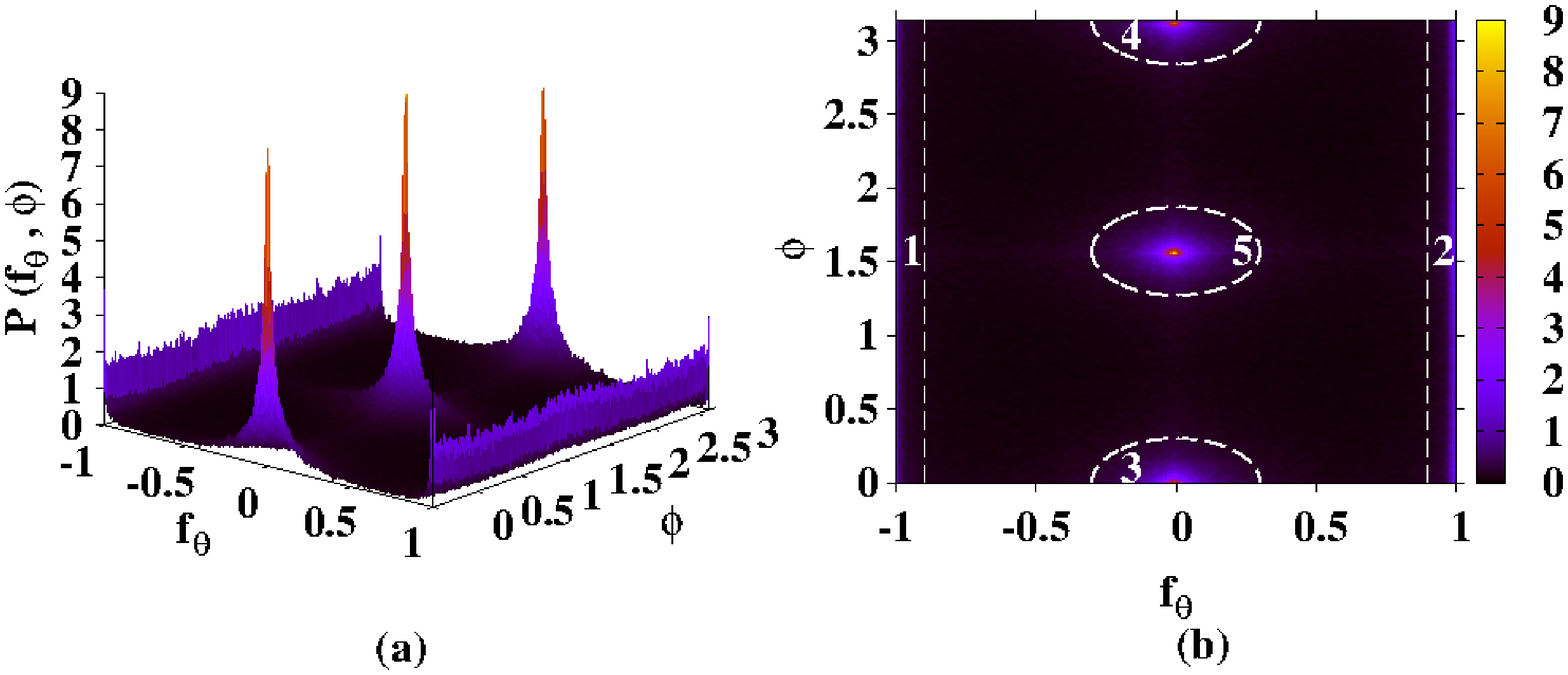}
 
 \includegraphics[width=0.8\textwidth]{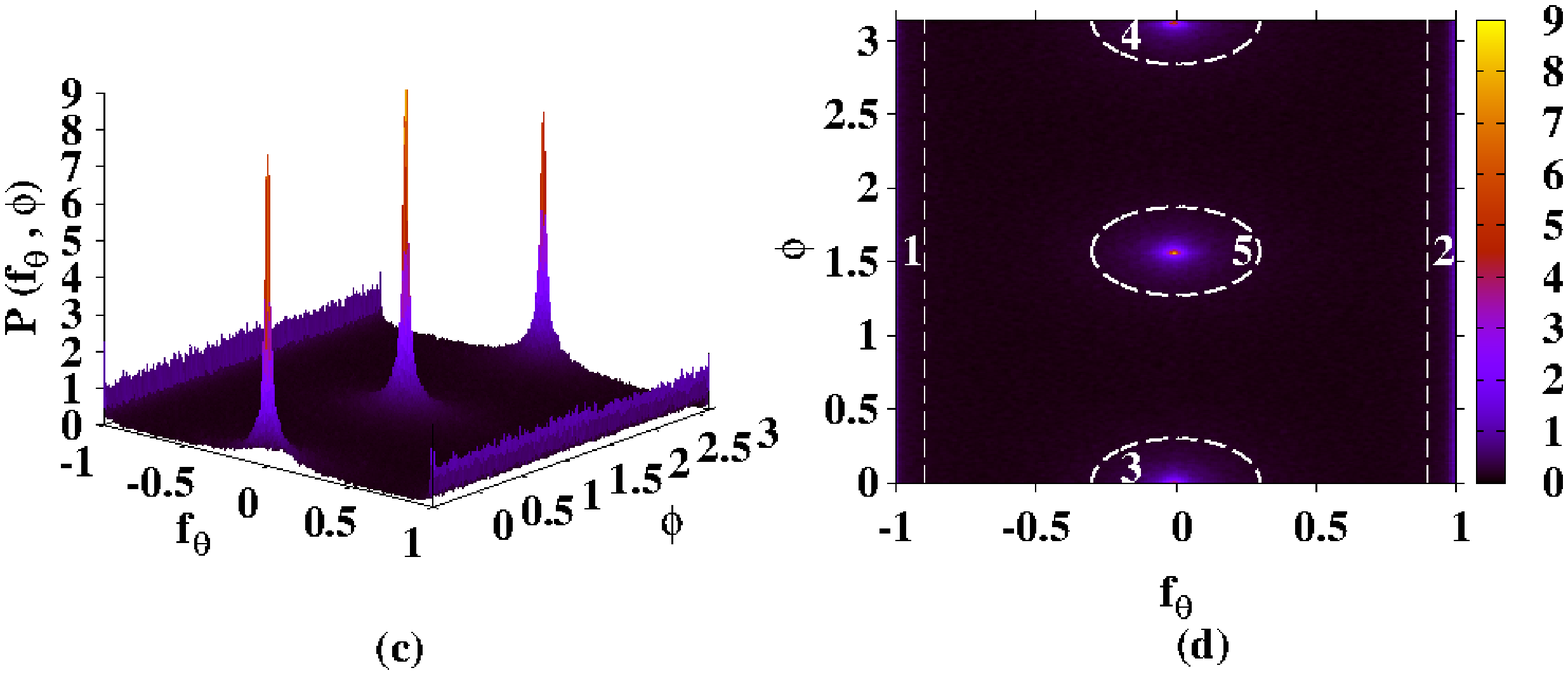}
 \caption{(Color online.) The probability distribution landscape, $P(f_{\theta},\phi)$, over the plane of $(f_{\theta},\phi)$, 
 in the case of a two-qubit state $\rho_{AB}$ of the form in Eq. (\ref{twoqubitequiv}) in the case of QD (a) and QWD (c). The regions 
 \textbf{1}--\textbf{5} are marked on the maps of the distribution landscape in the case of QD (b) and QWD (d) so that 
 the corresponding quantum correlation for majority of the states is optimized in the marked regions. The definition of the marked 
 regions are given in Eq. (\ref{marked}). The different shades in the figure correspond to different values of 
 $P(f_{\theta},\phi)$. All quantities plotted are dimensionless, except for $\phi$, which is in degrees. 
%  The $x$ axes of the figures represent $f_\theta=\cos \theta$, which is 
%  dimensionless. The $y$ axes of the figures denote $\phi$,
%  plotted in degrees, while $P(f_{\theta},\phi)$, plotted along the $z$ axes, is dimensionless.
 }
 \label{cor_par}
\end{figure*}

\subsubsection*{\textbf{Case 4: Triad}}
As the fourth scenario, we consider a very special earmarked set, called the ``triad'', where the set 
$\mathcal{S}_{E}$ consists of only the projection measurements corresponding to the three 
Pauli operators. This is an extremely restricted earmarked set, 
and one must expect large average VE if $Q_{c}$ is calculated for an arbitrary two-qubit state 
of rank $r$ by performing the optimization over the triad. However, in Sec. \ref{two_param}, 
we shall show that there exists a large class 
of two-qubit ``exceptional'' states for which the triad is equivalent to $\mathcal{S}_{C}$, with a vanishing VE. 

Before concluding the discussion on general two-qubit mixed states with different ranks, $r$, we briefly report the statistics of the VE, 
$\varepsilon_{n}^r$, where the subscript $n=3$ in the present case, denoting the size of the triad. 
We determine the probability, $P_{3}^r(\varepsilon_{3}^r)d\varepsilon_{3}^r$, that the value of $Q_{c}$, calculated for a 
randomly chosen two-qubit mixed state of rank $r$, has the VE between $\varepsilon_{n}^r$ and 
$\varepsilon_{3}^r+d\varepsilon_{3}^r$. To do so, as in the previous cases, 
we Haar uniformly generate $5\times10^{5}$ NPPT as well as PPT states for each value of $r=2,3$, and $4$. 
Fig. \ref{22err} depicts  the variations of normalized $P_{3}^r(\varepsilon_{3}^r)$  over the complete range $[0,1]$ of 
$\varepsilon_{n}^r$ for 
NPPT and PPT states of different ranks in the case of both QD as well as QWD. It is noteworthy that the distributions are sharply peaked 
in the low-error regions, and for a fixed rank $r$, the probability of finding a PPT state with a very low value 
of $\varepsilon_{3}^r$ is always higher than that for an NPPT state. 

\begin{figure}
 \includegraphics[scale=0.475]{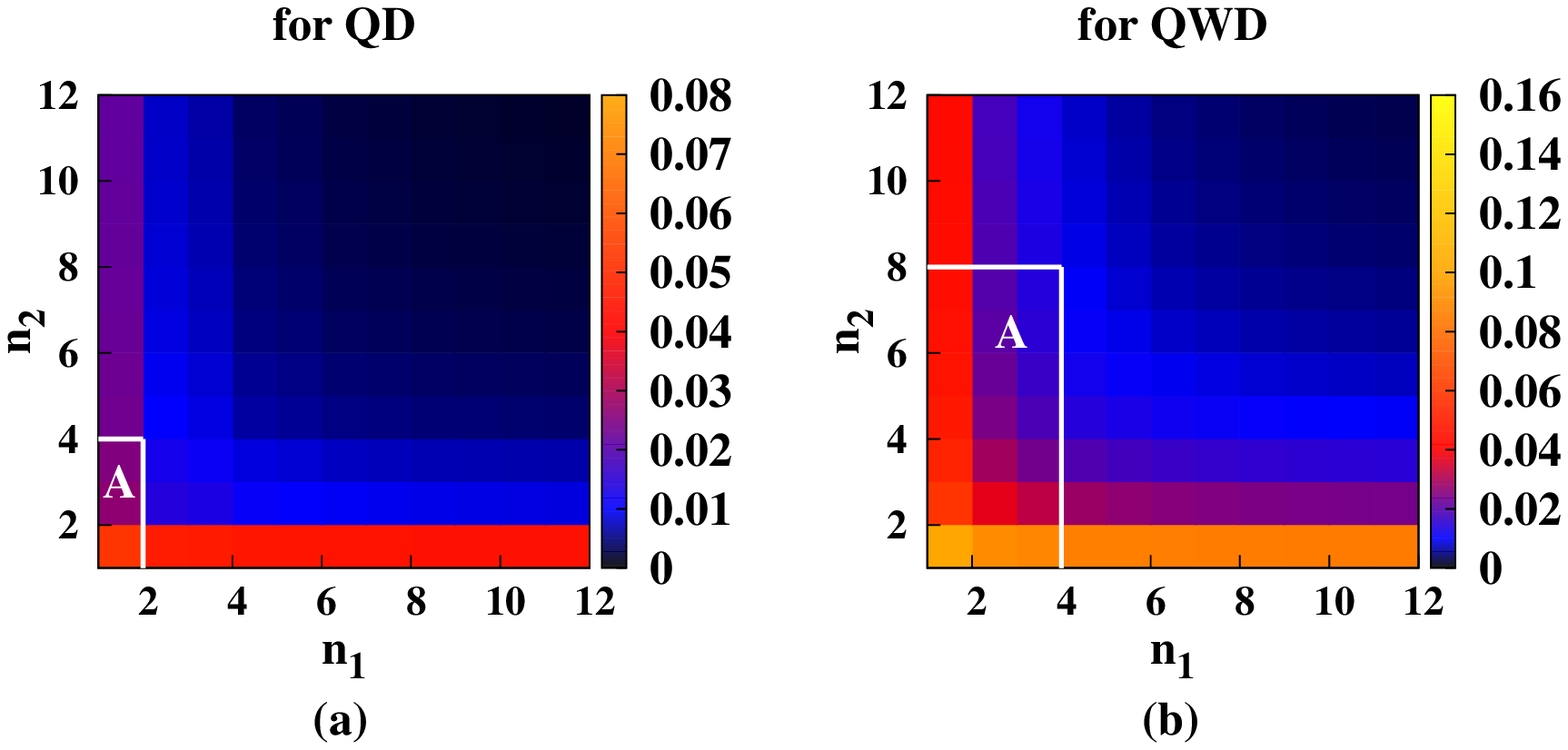}
 \caption{(Color online.) Variation of $\overline{\varepsilon}_{n}^{r}$ as a function of $n_{1}$ and $n_{2}$ ($n=n_{1}n_{2}$)
 in the case of two-qubit state, given in Eq. (\ref{twoqubitequiv}), where (a) QD and (b) QWD are considered as quantum 
 correlation measures. The ranges of $n_{1}$ and $n_{2}$ marked by $A$ are sufficient to obtain a considerably low value of 
 $\overline{\varepsilon}_{n}^{r}$ ($\sim10^{-3}$). The region is bigger in the case of QWD compared to that in QD.
 The different shades in the figure correspond to different values of $\overline{\varepsilon}_{n}^{r}$. All quantities plotted are dimensionless,
 except $\overline{\varepsilon}_{n}^r$ being in bits.}
 \label{surf2}
\end{figure}

\subsection{Two-qubit states in parameter space}
\label{two_param}

% Next, we consider the two-qubit states in the parameter space. 
A general two-qubit state, up to local unitary 
transformations \cite{luo_pra_2008}, can be written in terms of nine real parameters as 
\begin{eqnarray}
 \rho_{AB}&=&\frac{1}{4}[I_{A}\otimes I_{B}
 +\sum_{\alpha=x,y,z}c_{\alpha\alpha}\sigma_{A}^{\alpha}\otimes\sigma_{B}^{\alpha}\nonumber \\
 &&+\sum_{\alpha=x,y,z}c^{A}_{\alpha}\sigma_{A}^{\alpha}\otimes I_{B}
 +\sum_{\beta=x,y,z}c^{B}_{\beta}I_{A}\otimes \sigma_{B}^{\beta}].
 \label{twoqubitequiv}
\end{eqnarray}
Here, $c_{\alpha \alpha} = \langle\sigma^{\alpha} \otimes \sigma^{\alpha} \rangle$ are the 
\textquotedblleft classical\textquotedblright\; correlators given by the diagonal elements of the correlation matrix 
$(|c_{\alpha\alpha}|\leq1)$, 
$c^{A}_{\alpha} = \langle\sigma^{\alpha}_{A} \otimes I_{B}\rangle$ 
and $c^{B}_{\beta} = \langle I_{A} \otimes \sigma_{B}^{\beta} \rangle$ are the single site quantities called 
\textit{magnetizations} $(|c^A_{\alpha}|,|c^{B}_{\beta}|\leq1)$, 
given by the elements of the two local Bloch vectors, and
$I_{A}$ $(I_{B})$ is the identity operator on the Hilbert spaces of $A$ $(B)$.

The maximum rank of the two-qubit state given in Eq. (\ref{twoqubitequiv}) can be $4$. The probability distribution 
$P(f_{\theta},\phi)$, as defined in Eq. (\ref{para_pdf}), is obtained by generating $5\times10^5$  states of the form $\rho_{AB}$ by 
choosing the diagonal correlators and magnetizations randomly from their allowed ranges. Here, we drop the subscript $r$ for 
sake of simplicity. Fig. \ref{cor_par}(a) 
(for QD) and (c) (for QWD) 
show the profiles of $P(f_{\theta},\phi)$ which have three distinctly high populations around the set of values \textbf{(i)} 
($f_{\theta}=0,\phi=0,\pi$), 
\textbf{(ii)} ($f_{\theta}=0,\phi=\frac{\pi}{2}$), and \textbf{(iii)} ($f_{\theta}=\pm1,0\leq\phi\leq\pi$), which 
correspond to the eigenbasis of 
$\sigma^{x}$, $\sigma^{y}$, and $\sigma^{z}$, 
respectively, on the Bloch sphere. Let us now consider small regions around those three high density peaks. They are conveniently 
marked with numbers $1$--$5$ in Figs. \ref{cor_par}(b) and (d), and are defined as 
% 
% One can mark considerably small regions on the 
% probability distribution landscape such that the values of $(f_{\theta},\phi)$, for which the chosen quantum correlation for a large 
% fraction of states is optimized, belong to these regions. One such example is shown in Fig. \ref{cor_par}(b) (for QD) and 
% (c) (for QWD)). Here, the regions marked with numbers \textbf{1}--\textbf{5} are defined as
\begin{eqnarray}
 \mbox{\textbf{1}} &:& -1\leq f_{\theta}\leq -0.9,0\leq\phi\leq\pi, \nonumber \\
 \mbox{\textbf{2}} &:&  0.9\leq f_{\theta}\leq 1,0\leq\phi\leq\pi, \nonumber \\
 \mbox{\textbf{3}} &:& f_{\theta}^{2}+\phi^{2}\leq \omega^{2}, 0\leq\phi\leq\pi, \nonumber \\
 \mbox{\textbf{4}} &:& f_{\theta}^{2}+(\phi-\pi)^{2}\leq \omega^{2}, 0\leq\phi\leq\pi, \nonumber \\
 \mbox{\textbf{5}} &:& f_{\theta}^{2}+(\phi-\frac{\pi}{2})^{2}\leq \omega^{2}, 0\leq\phi\leq\pi, 
 \label{marked}
\end{eqnarray}
with $\omega=0.3$. In the case of QD, about $56.64\%$ of the total number of sample states, given in Eq. \ref{twoqubitequiv}, 
are optimized in the region marked in 
Fig. \ref{cor_par}(b), while the percentage is approximately $42.2\%$ in the case of QWD (Fig. \ref{cor_par}(d)). Note that these are 
considerably high fractions, taking into account the fact that the area of the marked regions combined together is small 
compared to the entire area of the parameter space. 
% Considering these numerical results, it is, therefore, reasonable to study the 
% scaling of VE starting from the triad as the earmarked set, and increasing the size of the set by considering the projectors defined 
% by the parameters in the marked regions.

We also consider the optimization of QD and QWD in the case of the two-qubit state given in Eq. (\ref{twoqubitequiv}) by 
confining the earmarked set to a collection of projection measurements corresponding to a uniformly distributed set of 
points on the entire surface of the Bloch sphere. 
As discussed in Sec. \ref{gen_mix}, we divide the 
ranges of the parameters $\phi$ and $f_{\theta}$ by $n_{1}$ and $n_{2}$ equispaced intervals, and perform the minimization of 
QD and QWD over the set $\mathcal{S}_E$ of size $n=n_{1}n_{2}$. We find that a significantly low value of average VE is achieved 
in the case of QD when $n_{1}\geq2$ and $n_{2}\geq4$. However, as observed in all the previous cases, the required size  
of the earmarked set, $\mathcal{S}_E$, to obtain an average VE of same order as in QD, is larger $(n_{1}\geq4,n_{2}\geq8)$ when QWD
is taken as the measure of 
quantum correlation, as clearly visible in Fig. \ref{surf2}.

\subsubsection*{\textbf{Special case: Mixed states with fixed magnetizations}} 

We conclude the section by discussing a special case of the two-qubit state given in Eq. (\ref{twoqubitequiv}),
where apart from the diagonal correlators, any one of the three magnetizations is non-zero while the other two magnetizations 
vanish. In particular, we consider a two-qubit state of the form
\begin{eqnarray}
\rho_{m}&=&\frac{1}{4}(I_{A}\otimes I_{B}+\sum_{\alpha=x,y,z}c^{\alpha\alpha}\sigma_{A}^{\alpha}\otimes\sigma_{B}^{\alpha}\nonumber\\ 
&&+m_{A}^{\beta}\sigma_{A}^{\beta}\otimes I_{B}+m_{B}^{\beta}I_{A}\otimes\sigma_{B}^{\beta}),
\label{rhom}
\end{eqnarray} 
with $\beta=x,y$, or $z$.
Here, $m_{i}^{\beta}=\mbox{tr}(\sigma^{\beta}_{i}\rho_{i})$ $(\beta=x,y,z)$ is the magnetization with $\rho_{i}$ being the local density 
matrix of the qubit $i$ ($i=A,B$).  The two-qubit $X$-state \cite{xstate} of the form  
\begin{equation}
\rho_{AB}^{X}=
\left(
\begin{array}{cccc}
   a_{1}&0&0&b_{1}\\
   0&a_{2}&b_{2}&0\\
   0&b_{2}&a_{3}&0\\
   b_{1}&0&0&a_{4}
\end{array}
\right),
\label{X}
\end{equation}
written in the computational basis $\{|00\rangle,|11\rangle,|01\rangle,|10\rangle\}$
is a special case of $\rho_{m}$ with $\beta=z$. The matrix elements, 
$\{a_{i}:i=1,\cdots,4\}$ and $\{b_{j}:j=1,2\}$, are real numbers,
and can be considered as functions of the correlators $c_{\alpha\alpha}$ $(\alpha=x,y,z)$, 
and the magnetizations, $m_{A}^{z}$ and $m_{B}^{z}$. 
The importance of two-qubit states of the form given in Eq. (\ref{X}) includes the fact that they are found to occur in the 
quantum information theoretic analyses of several well-known quantum spin systems, such as the one-dimensional $XY$ model 
in a transverse field \cite{xy_group}, and the $XXZ$ model \cite{xxz_group}.  
These models possess certain symmetries that govern the two-qubit reduced density matrices obtained by tracing out all 
other spins except two chosen spins from their ground, thermal, as well as time-evolved states with specific time 
dependence, to have the form given in Eq. (\ref{X}) \cite{modi_rmp_2012,rev_xy,ent_xy_bp,xy_discord}. 
Since completely analytical forms of QD and QWD are not yet available \cite{disc_2q}, numerical techniques need to be 
employed, and our methodology show a path to handle them analytically. See \cite{freezing_paper,disc_num_anal} 
in this respect.

For the purpose of demonstration, let us assume that the magnetization of $\rho_{m}$ is along the $x$ direction. Generating a
large number $(5\times10^5)$ of such
states by randomly choosing the correlators and the magnetizations within their allowed ranges of values, and performing 
extensive numerical analysis, we find that for about $99.97\%$ of the states, it is enough to perform the optimization over 
the triad ($\mathcal{S}_{E}$ consists of the projection measurements corresponding to the three Pauli spin operators) 
to determine actual value of QD. 
This is due to the fact that these $99.97\%$ of states are ``exceptional'', i.e., the VE, $\varepsilon_{3}=0$ for all these states when 
the optimization is performed over $\mathcal{S}_E$ (as discussed in Sec. \ref{def_err}). Here, $\varepsilon_{3}$ is the VE where we 
drop the superscript $r$ for simplicity.  
% for which the optimization of QD occurs for a 
% local projector $\Pi_{opt}^{A}\in\mathcal{S}_{E}$.
In the case of the remaining $0.03\%$ of states, for which the optimal projection measurement does not belong to 
$\mathcal{S}_{E}$. The VE, resulting from the optimization performed over $\mathcal{S}_E$, is found to be
$\varepsilon_{3}\leq2.9088\times10^{-3}$. Similar result has been reported in \cite{num-err_group}, although 
our numerical findings result in a different bound. As an example, we consider the state
$\rho_{m}$ defined by $c^{xx}=(-1)^n\times0.956861$, $c^{yy}$ (or $c^{zz}$) $=(-1)^m\times0.267575$, $c^{zz}$ 
(or $c^{yy}$) $=(-1)^{n+m+1}\times0.275867$,
$m_{A}^{x}=(-1)^p\times0.94976$, and $m_{B}^{x}=(-1)^p\times0.907559$ with $m$, $n$, and $p$ being integers, even or odd. 
There exists eight such states (corresponding to different values of $m$, $n$, and $p$) 
for which $\varepsilon_{3}=2.9088\times10^{-3}$. 
Amongst the set of exceptional states, the optimal projector is $\sigma^{x}_A$ for about $27.4\%$ states. 
For the rest of the set of exceptional states, QD of the half of them (i.e., $36.3\%$ states) are optimized for 
with projection measurement corresponding to $\sigma^{y}_A$ 
while the rest are optimized for the projector $\Pi_{opt}^{A}$ corresponding to $\sigma^{z}_A$. 

Interestingly, all of the above results except one remains invariant with 
a change in the direction of magnetization. For example, a change in the direction of magnetization from $x$ to $z$ 
results in the optimization of $27.4\%$ of the set of exceptional states for projection measurement corresponding to 
$\sigma_{A}^{z}$. This highlights an underlying symmetry of 
the state suggesting that the presence of magnetization $\langle\sigma^{\alpha}\rangle$ diminishes the 
probability of optimization of QD for $\Pi_{opt}^{A}$ corresponding to $\sigma_{A}^{\alpha}$, $\alpha=x,y,z$, 
provided the state $\rho_{m}$ belongs to the set of exceptional states.

Note also that our approach offers a closed form expression for CQD in the case of two-qubit states of the form 
given in Eq. (\ref{X}) as 
\begin{eqnarray}
 D_c=S(\rho_A)
%  -\sum_{i=1}^2\alpha_i\log_2\alpha_i+\sum_{i=1}^{4}\beta_i\log_2\beta_i
-S(\rho_{AB}^X)+\min\Big[S^\prime,S^\prime_{\pm}\Big],
 \label{xdisc}
\end{eqnarray}
where 
% $S(\rho_A)$ $(S(\rho_{AB}^X))$ is the von Neumann entropy of the local density matrix, 
$\rho_A=\mbox{Tr}_{B}[\rho_{AB}^X]$. 
% $(\rho_{AB}^{X})$. 
% with $\alpha_1=a_1+a_2$, $\alpha_2=a_3+a_4$, and 
% \begin{eqnarray}
%  \beta_1&=&\frac{1}{2}\left[a_1+a_4+\sqrt{(a_1-a_4)^2+4b_1}\right]\nonumber \\
%  \beta_2&=&\frac{1}{2}\left[a_1+a_4-\sqrt{(a_1-a_4)^2+4b_1}\right]\nonumber \\
%  \beta_3&=&\frac{1}{2}\left[a_2+a_3+\sqrt{(a_2-a_3)^2+4b_2}\right]\nonumber \\
%  \beta_4&=&\frac{1}{2}\left[a_2+a_3+\sqrt{(a_2-a_3)^2+4b_2}\right].
%  \label{betavalues}
% \end{eqnarray}
The quantities $S^\prime$ and $S^\prime_{\pm}$ are functions of the matrix elements $\{a_i;i=1,\cdots,4\}$, 
and $\{b_j;j=1,2\}$, and are given by 
\begin{eqnarray}
 S^\prime&=&(a_1+a_2)\log_2(a_1+a_2)+(a_3+a_4)\log_2(a_3+a_3)\nonumber\\
 &&-\sum_{i=1}^4a_i\log_2 a_i,\\
 S^\prime_{\pm}&=&1-\frac{1}{2}\sum_{i=1}^2\alpha^{\pm}_i\log_2\alpha^{\pm}_i,
\end{eqnarray}
where 
\begin{eqnarray}
 \alpha^{\pm}_i=1+(-1)^i\sqrt{(a_1-a_2+a_3-a_4)^2+4(b_1\pm b_2)^2}.\nonumber\\
 \label{alphapm}
\end{eqnarray}
% and
% \begin{eqnarray}
%  {\xi_{\pm}}^2=(a_1-a_2+a_3-a_4)^2+4(b_1\pm b_2)^2.
%  \label{xipm}
% \end{eqnarray}
The expression of $D_c$ given in Eq. (\ref{xdisc}) is exact for the 
exceptional states, and results in a very small absolute error in the case of all other two-qubit $X$ states,
when measurement over qubit $A$ is considered.

The numerical results for the state $\rho_{m}$ depends strongly on the choice of quantum correlation measures. To demonstrate this, 
we compute QWD instead of QD for $\rho_{m}$. Our numerical analysis suggests 
that irrespective of the direction of magnetization, for about $95.51\%$ of the two-qubit states of the form $\rho_{m}$, the QWD is 
optimized over the triad $\mathcal{S}_{E}$, resulting $\varepsilon_3=0$. 
These states constitute the set of exceptional states in the case of QWD.
For the rest $4.49\%$ of states, the assumption that $\Pi_{opt}^{A}\in\mathcal{S}_{E}$ results in a VE, 
$\varepsilon_{3}=W_{c}-W_{a}\leq1.0076\times10^{-1}$, where $W_{c}$ is the CQWD, and $W_{a}$ is the actual value of QWD. 
Note that the upper bound of the absolute error, in the case of the QWD, is much higher
than that for QD of $\rho_{m}$. In contrast to the case of the QD, it is observed that for a 
state $\rho_{m}$ with magnetization along, say, the $x$ direction, within the set of exceptional states, the QWD for 
$\geq53\%$ of states -- a larger percentage -- is optimized for $\Pi_{opt}^{A}$ corresponding to $\sigma_{A}^x$. 
The rest of the exceptional states, 
with respect to optimization of QWD, are equally distributed over the cases where $\Pi_{opt}^{A}$ corresponds to 
$\sigma_{A}^y$ and $\sigma_{A}^z$.
Similar to the case of QD (Eq. (\ref{xdisc})), 
one can obtain a closed form expression of the CQWD in the case of two-qubit $X$ states 
as 
\begin{eqnarray}
 W_c=\min\left[\tilde{S},\tilde{S}_{\pm}\right]-S(\rho_{AB}^X),
 \label{xdef}
\end{eqnarray} 
where the quantities $\tilde{S}$ and $\tilde{S}_{\pm}$ are given by $\tilde{S}=-\sum_{i=1}^4 a_{i}\log_2 a_{i}$, 
and $\tilde{S}_{\pm}=-\sum_{i=1}^4 (\alpha^{\pm}_{i}/4)\log_2(\alpha^{\pm}_{i}/4)$ respectively, 
with $\alpha^{\pm}_i$ given in Eq. (\ref{alphapm}).
For the $95.51\%$ of states with the form of $X$ states, $W_c$ provides a closed form for QWD, when measurement 
is done on qubit $A$.

\subsection{Application: Quantum spin systems}
\label{apply}

Now we discuss how the constrained optimization technique introduced in the paper, and its potential to provide 
closed form expresions of quantum correlations with small errors, can help in analyzing physical systems. 
For the purpose of demonstration, we choose the spin-$\frac{1}{2}$
anisotropic quantum XY model in an external transverse field, which has been studied extensively using 
quantum information theoretic measures \cite{modi_rmp_2012,rev_xy,ent_xy_bp,xy_discord}. Successful laboratory implementation 
of this model in different substrates \cite{solid_xy,num_trap_xy,trap_rev,lab_xy} has allowed experimental
verification of properties of several measures of quantum correlations leading to a better
understanding of the novel properties of the model. In this paper, we specifically consider 
variants of the model, viz.,
\textbf{(i)} the zero-temperature scenario where the model is defined on a lattice of $N$ sites, 
with an external homogeneous transverse field, 
and \textbf{(ii)} the two-qubit representation of the model with staggered transverse field, at finite temperature.

\subsubsection*{\texorpdfstring{$L$}{\textit{\textbf{L}}}-\textbf{qubit system with homogeneous field}}

The Hamiltonian describing the anisotropic $XY$ model in an external homogeneous transverse field \cite{xy_group}, 
with periodic boundary condition, is given by \cite{xy_group,rev_xy}
\begin{eqnarray}
H&=&\frac{J}{2}\sum_{i=1}^{L}
\left\{(1+g)\sigma_{i}^{x}\sigma_{i+1}^{x}+(1-g)\sigma_{i}^{y}\sigma_{i+1}^{y}\right\}
+h\sum_{i=1}^L\sigma_{i}^{z},\nonumber\\
\label{xy}
\end{eqnarray}
where $J$, $g$ ($-1\leq g\leq1$), and $h$  are the coupling strength, the anisotropy parameter, and the strength of
the external homogeneous transverse magnetic field, respectively. At zero temperature and in the thermodynamic limit
$L\rightarrow\infty$, the ground state
of the model encounters a quantum phase transition (QPT) \cite{qpt_book} at $\lambda=\lambda_c=1$ $(\lambda=J/h)$
from a quantum paramagnetic phase to an antiferomagnetic phase \cite{xy_group,rev_xy,qpt_book}. A special 
case of the model is given by the well-known transverse-field Ising model $(g=1)$.
The Hamiltonian, $H$, can be exactly diagonalized by the successive applications of the Jordan-Wigner and the Bogoliubov 
transformations, and the single-site magnetization, $m_i^z$, and the two-spin correlation functions, 
$c^{\alpha\alpha}_{ij}$, of the spins $i$ and $j$ ($i,j\in\{1,2,\cdots,L\},i\neq j$, and $\alpha=x,y,z$) 
can be determined. Using these 
parameters, one can obtain the two-spin reduced density matrix, $\rho_{ij}$, for the ground state of the model,
which is of the form given in Eq. (\ref{X}), where the matrix elements are functions of the single-site
magnetization, and two-site spin correlation functions. 
% where the matrix elements are given by
% \begin{eqnarray}
%  a_1&=&\frac{1}{4}\Big(1+m_i^z+m_j^z+c^{zz}_{ij}\big),\nonumber\\
%  a_2&=&\frac{1}{4}\Big(1+m_i^z-m_j^z-c^{zz}_{ij}\Big),\nonumber\\
%  a_3&=&\frac{1}{4}\Big(1-m_i^z+m_j^z-c^{zz}_{ij}\Big),\nonumber\\
%  a_4&=&\frac{1}{4}\Big(1-m_i^z-m_j^z+c^{zz}_{ij}\Big),\nonumber\\
%  b_1&=&\frac{1}{4}\Big(c^{xx}_{ij}+c^{yy}_{ij}\Big),\,
%  b_2=\frac{1}{4}\Big(c^{xx}_{ij}-c^{yy}_{ij}\Big),
% \end{eqnarray}
% where $m^z_i=m^z_j$ due to translational invariance of the model. 

For a finite sized spin-chain, the QPT at $\lambda_c=1$ for $|g|<1$ is detected by a maximum in the variation of
$\frac{dQ}{d\lambda}$ against the tuning parameter, $\lambda$, which occurs in the vicinity of $\lambda_c=1$.
Here, $Q$ is the measure of quantum correlation, viz., QD and QWD, computed for the nearest-neighbour ($|i-j|=1$)
reduced density matrix $\rho_{ij}$, obtained from the ground state of the model.
With the increase in system size, the maximum sharpens and the QPT point approaches $\lambda_c=1$ as 
$\lambda^L_c=\lambda_c+\alpha L^{-\gamma}$, where $\lambda^L_c$ is the value of $\lambda$ at which the maximum of 
$\frac{dQ}{d\lambda}$ occurs for a fixed value of $L$, $\alpha$ is a dimensionless constant, and $\gamma$ is the 
scaling index.

\begin{figure}
 \includegraphics[scale=0.6]{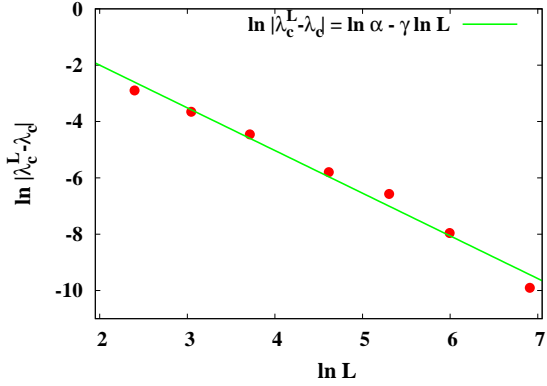}
 \caption{(Color online.) Finite size scaling analysis for the transverse-field $XY$ model 
 using CQWD, computed from Eq. (\ref{xdef}), as the observable. 
 The QPT point for a system of size $L$ approaches $\lambda_c=1$ as 
 $\lambda_c^L=\lambda_c+\alpha L^{-\gamma}$, where $\gamma=1.515$ and $\alpha=1.031$. 
 All quantities plotted are dimensionless. The abscissa of the figure is in natural logarithm of the number of qubits, while the 
 ordinate is in natural logarithm of $\lambda$, a dimensionless quantity.}
 \label{xyscaling}
\end{figure}

We perform the scaling analysis in the case of the transverse-field $XY$ model by using CQD and CQWD as observables, and 
computing their values using Eqs. (\ref{xdisc}) and (\ref{xdef}), respectively. 
Fixing the value of the anisotropy parameter at $g=0.5$, using CQD, the scaling parameters are obtained as 
$\alpha=0.109$ and $\gamma=1.215$, while scaling analysis using the CQWD results in 
$\alpha=1.031$, and $\gamma=1.515$. This indicates a higher value of $\gamma$ in the case of CQWD, and 
therefore a better finite-size scaling in comparison to CQD.
These scaling parameters are consistent with the scaling parameters obtained by performing the finite-size scaling 
analysis using unconstrained optimization for computing QD and QWD numerically.  
This indicates that 
the CQWD can capture the finite-size scaling features perfectly for the transverse-field $XY$ model in the vicinity of the 
QPT. Therefore, our methodology provides a path to explore the 
quantum cooperative phenomena occuring in quantum spin models, in terms of different measures of quantum correlations 
that involve an optimization, in a numerically beneficial way, or analytically.
Fig. \ref{xyscaling} provides the log-log plot of the variation of $|\lambda^L_c-\lambda_c|$ with $L$, 
where $W_c$ (Eq. (\ref{xdef})) is used.

\subsubsection*{\textbf{Two-qubit system with inhomogeneous transverse field}}

We now study the two-qubit anisotropic $XY$ model in the presence of staggered transverse magnetic field, represented by 
the Hamiltonian 
\begin{eqnarray}
H_2&=&J\left\{(1+g)\sigma_{1}^{x}\sigma_{2}^{x}+(1-g)\sigma_{1}^{y}\sigma_{2}^{y}\right\}
+\sum_{i=1}^2h_i\sigma_{i}^{z},\nonumber\\
\label{xy2} 
\end{eqnarray}
where the strength of the external field on qubit $i$ is given by $h_i$ ($i=1,2$), and all the other 
symbols have their usual meaning. The thermal state of the two-qubit system, $\rho_T$, at a temperature 
$T$, is given by 
$\rho_T=\sum_{i=0}^3e^{-\beta E_j}P[|\psi_j\rangle]/Z$, where the eigenenergies $\{E_j\}$, 
and the eigenvectors $\{|\psi_j\rangle\}$, $j=0,\cdots,3$ are obtained by diagonalizing the Hamiltonian (Eq. (\ref{xy2})),
$\beta=\frac{1}{k_BT}$ with $k_B$ being the Boltzman constant, 
and $P[|\alpha\rangle]=|\alpha\rangle\langle\alpha|$. Here, $Z=\mbox{tr}[e^{-\beta H_2}]$ is the partition function 
of the system. Similar to the previous example, the form of $\rho_T$ is as given in Eq. (\ref{X}), where 
the matrix elements are given by 
\begin{eqnarray}
 a_1&=&\frac{1}{2h_+^2u}\Big[h_+^2\cosh(\beta h_+)-h_+(h_+^2-4g^2)^\frac{1}{2}\sinh(\beta h_+)\Big],\nonumber \\
 a_2&=&\frac{1}{2h_-^2u}\Big[h_-^2\cosh(\beta h_-)+h_-(h_-^2-4)^\frac{1}{2}\sinh(\beta h_-)\Big],\nonumber \\
 a_3&=&\frac{1}{2h_-^2u}\Big[h_-^2\cosh(\beta h_-)-h_-(h_-^2-4)^\frac{1}{2}\sinh(\beta h_-)\Big],\nonumber \\
 a_4&=&\frac{1}{2h_+^2u}\Big[h_+^2\cosh(\beta h_+)+h_+(h_+^2-4g^2)^\frac{1}{2}\sinh(\beta h_+)\Big],\nonumber \\
 b_1&=&-\frac{g\sinh(\beta h_+)}{h_+u},\,
 b_2=-\frac{\sinh(\beta h_-)}{h_-u},
 \label{mat_el}
\end{eqnarray}
with $h_+=\Big[4g^2+\frac{(h_1+h_2)^2}{J^2}\Big]^\frac{1}{2}$, $h_-=\Big[4+\frac{(h_2-h_1)^2}{J^2}\Big]^\frac{1}{2}$, and 
$u=\cosh(\beta h_+)+\cosh(\beta h_-)$. Using Eqs. (\ref{xdisc}), (\ref{xdef}), and (\ref{mat_el}), one can 
determine CQD and CQWD for the thermal state of the model as functions of the system parameters, $g$, $\frac{h_1}{J}$, 
and $\frac{h_2}{J}$, 
and the temperature $T$. 

\begin{figure}
 \includegraphics[scale=0.345]{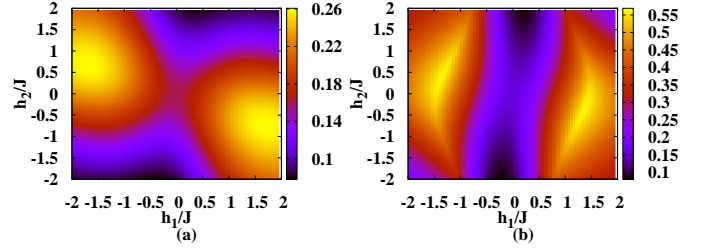}
 \caption{(Color online.) Variations of (a) CQD and (b) CQWD, as computed from Eqs. (\ref{xdisc}) and (\ref{xdef}), 
 respectively, against $\frac{h_1}{J}$ and $\frac{h_2}{J}$, with $g=0.5$, for the two-qubit system with 
 inhomogeneous transverse field at a finite temperature, given by $J\beta=1$. All quantities plotted are dimensionless, except for 
 CQD and CQWD, which are in bits.}
 \label{2qubit}
\end{figure}

The variations of CQD and CQWD with $\frac{h_1}{J}$ and $\frac{h_2}{J}$, as calculated using Eqs. (\ref{xdisc}) and 
(\ref{xdef}),
for $g=0.5$ and $J\beta=1$, are plotted in Fig. \ref{2qubit}. The analytical form in Eq. (\ref{xdisc}) is exact
for QD in the present case, while Eq. (\ref{xdef}) results in a maximum VE, $\varepsilon_{max}=6.26\times10^{-2}$
in the values of QWD, occurring at the points $(\frac{h_1}{J},\frac{h_2}{J})=(\pm1.45,\pm0.55)$, 
for the given ranges of the system parameters, viz.,
$\left|\frac{h_1}{J}\right|,\left|\frac{h_2}{J}\right|\leq2$. However, the qualitative features of the 
variation of QWD with $\frac{h_1}{J}$ and $\frac{h_2}{J}$, 
when computed using Eq. (\ref{xdef}), are similar to those when the QWD is computed via unconstrained optimization. 
Also, if one adopts the constrained optimization technique discussed in Case 3 of Sec. \ref{gen_mix}, $\varepsilon_{max}$
reduces to $\sim 10^{-6}$ for $n_1=8$, and $n_2=1$. Hence, the value of QWD, with negligible error, can be obtained in the 
case of the thermal state of the two-qubit anisotropic $XY$ model in an external inhomogeneous magnetic field with very 
small computational effort, if the constraints are used appropriately. This again proves the usefulness of our methodology. 

\section{Restricted quantum correlations for bound entangled states}
\label{bestates}

From the results discussed in the previous section, 
it comes as a common observation that the VE is less in the case of two-qubit PPT states when compared to the two-qubit NPPT states
of fixed rank for a fixed measure of quantum correlation.  
A natural question arising out of the previous 
discussions is whether the result of low VE in the case of PPT states, which until now were all separable states, 
holds even when the state is entangled.
Since PPT states, if entangled, are always BE, to answer this question, one has to look beyond $\mathbb{C}^{2}\otimes\mathbb{C}^{2}$ 
systems and consider quantum states in higher dimensions where PPT bound entangled states exist. 
However, generating BE states in higher dimension is itself a non-trivial problem. 
Instead, we focus on a number of paradigmatic BE states in $\mathbb{C}^{2}\otimes\mathbb{C}^{4}$ and 
$\mathbb{C}^{3}\otimes\mathbb{C}^{3}$ systems and investigate the properties of the VE in a case-by-case basis.

It is also observed, from the results reported in the previous section, that the choice of the triad as the earmarked set yields 
good result in the context of low VE in the case of a large fraction of two-qubit states in the parameter space. 
Motivated by this observation, in the following calculations, we choose the triad 
constituted of projection measurements corresponding to the three spin-operators, $\{S^{x},S^{y},S^{z}\}$, 
in the respective physical system denoted by $\mathbb{C}^d$,  
as the earmarked set, $\mathcal{S}_E$. Here, $S^{\beta}$, $\beta=x,y,z$, 
are the spin operators for a spin-$\frac{d-1}{2}$ particle (a system of 
dimension $d$).
% Note that in the case of $d=2$, the triad is defined by $\mathcal{S}_{E}=\{\sigma^{x},\sigma^{y},\sigma^{z}\}$, as described in the 
% previous section. 
From now on, in the case of the triad as the earmarked set, we discard the subscript `3', and denote the VE by $\varepsilon$ for 
sake of simplicity. If the dimension is $2$, the earmarked set is given by the triad 
constituted of the projection measurements corresponding to  three Pauli matrices.

\begin{figure}%[h]
 \includegraphics[scale=0.675]{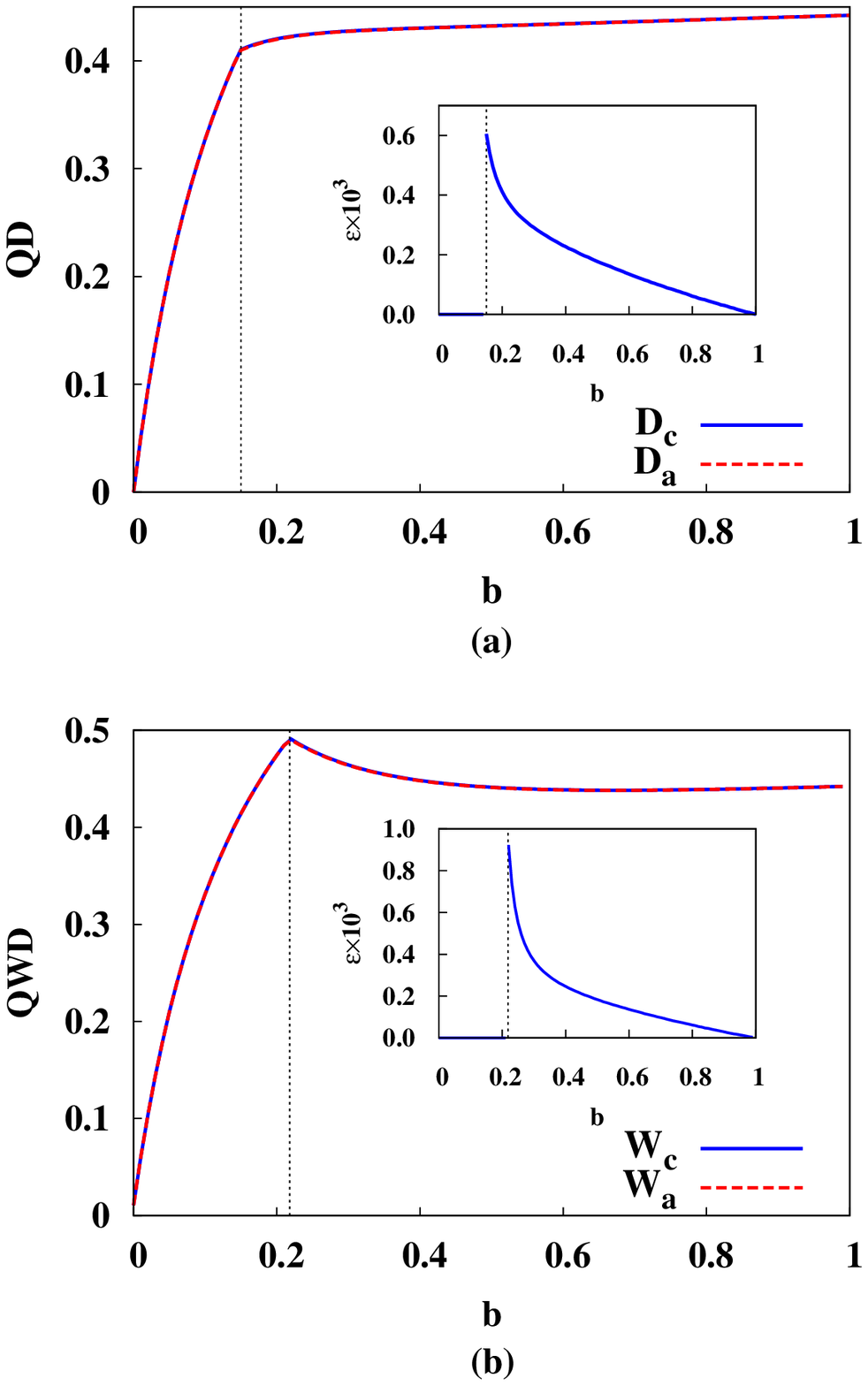}
 \caption{(Color online.) Variation of QD (a) and QWD (b) as a function of $b$ in the case of the PPT BE state $\rho_{b}$ in 
 $\mathbb{C}^{2}\otimes\mathbb{C}^{4}$ systems. (Inset) Variation of the corresponding VE, $\varepsilon$, as a function of 
 $b$. The error jumps from zero to a non-zero value at $b=0.15$ in the case of QD, while at $b=0.22$ in the case of QWD. 
 The quantities $D_c$, $D_a$, $W_c$, $W_a$, and $\varepsilon$ are in bits, while $b$ is dimensionless.}
 \label{24be1}
\end{figure}

\subsection{\texorpdfstring{$\mathbb{C}^{2}\otimes\mathbb{C}^{4}$}{} system}
\label{2x4system}

The first PPT BE state that we consider is in a $\mathbb{C}^{2}\otimes\mathbb{C}^{4}$ system, and is given by \cite{be1}
\begin{eqnarray}
 \rho_{b}=
\frac{1}{7b+1}\left(
 \begin{array}{cccccccc}
  b & 0 & 0 & 0 & 0 & b & 0 & 0 \\
  0 & b & 0 & 0 & 0 & 0 & b & 0 \\
  0 & 0 & b & 0 & 0 & 0 & 0 & b \\
  0 & 0 & 0 & b & 0 & 0 & 0 & 0 \\
  0 & 0 & 0 & 0 & f_{b} & 0 & 0 & g_{b} \\
  b & 0 & 0 & 0 & 0 & b & 0 & 0 \\
  0 & b & 0 & 0 & 0 & 0 & b & 0 \\
  0 & 0 & b & 0 & g_{b} & 0 & 0 & f_{b}
 \end{array}
\right),
\label{2x4state}
\end{eqnarray}
with 
\begin{eqnarray}
f_{b}=\frac{1+b}{2},\,g_{b}=\frac{\sqrt{1-b^{2}}}{2}, 
\label{fbgb}
\end{eqnarray}
and $0\leq b\leq 1$. The state is BE for all allowed values of $b$. 
We calculate QD and QWD of the state by performing measurement on the qubit.
% whose Hilbert space is spanned by the computational basis 
% $\{|0\rangle,|1\rangle\}$, i.e., the eigenbasis of $\sigma_{z}$. 
% which is similar to the two-qubit case. 
In Fig. \ref{24be1}(a), we plot the CQD, $D_{c}$, and the actual QD, $D_{a}$,
as functions of the state parameter, $b$. At $b=0.15$, QD shows a sudden change in its variation. 
The inset of Fig. \ref{24be1}(a) shows the variation of the relative error, $\varepsilon$, with $b$, exhibiting a discontinuous 
jump from zero to a non-zero value at $b=0.15$. Note that for $b<0.15$, $\varepsilon=0$, indicating $\Pi_{opt}^{A}\in\mathcal{S}_E$, 
which, in this case, corresponds to $\sigma^{z}$. For $b\geq 0.15$, 
the maximum value of $\epsilon$, although non-zero, is of the order of $10^{-3}$, and monotonically 
decreases to zero at $b=1$. Therefore, a restriction of the minimization of QD over the triad, in this region, 
is advantageous. For $b\geq0.15$, we observe that minimization in obtaining the value of $D_{c}$ is achieved from 
the projection measurement corresponding to $\sigma^{x}$. 
Using this information, one can obtain a closed form expression for 
$D_c$ as 
\begin{eqnarray}
D_c=S(\rho_{b}^A)
-S(\rho_{b})+\min\Big[\bar{S}_1,\bar{S}_2\Big],
 \label{24disc}
\end{eqnarray}
where $S(\rho_b^A)$ is the von Neumann entropy of the local density matrix of the qubit part.  The quantities 
$\bar{S}_1$ and $\bar{S}_2$ are functions of the state parameter, $b$, given by
\begin{eqnarray}
 \bar{S}_1&=&\frac{1}{1+7b}\Big[1+9b+(1+3b)\log_2(1+3b)-2b\log_2b\nonumber \\
          &&-\frac{1}{2}\sum_{i=1}^2\zeta_i\log_2\zeta_i\Big], \\
 \bar{S}_2&=&-\frac{1}{2}\sum_{i,j=1}^{2}(\tau_{ij}\log_2\tau_{ij}+\tau^\prime_{ij}\log_{2}\tau^\prime_{ij}),
\end{eqnarray}
where the quantities $\zeta_i$, $\tau_i$, and $\tau_i^\prime$ are given by 
\begin{eqnarray}
\zeta_i&=&1+b+(-1)^i\sqrt{1-b^2},\nonumber \\
\tau_{ij}&=&\frac{1}{4(1+7b)}\Big(1+9b+(-1)^i\sqrt{1-x^2}+(-1)^j\omega_{i}\Big),\nonumber \\
\tau^\prime_{ij}&=&\frac{1}{4(1+7b)}\Big(1+5b+(-1)^i\sqrt{1-x^2}+(-1)^j\omega^\prime_{i}\Big),\nonumber \\
\label{quant}
\end{eqnarray}
with $\omega_{i}^2=2\left[1-3b+12b^2+(-1)^i(1-3b)\sqrt{1-b^2}\right]$, and
${\omega^\prime_{i}}^2=2\left[1+b+8b^2+(-1)^i(1+b)\sqrt{1-b^2}\right]$.
The expression in Eq. (\ref{24disc}) provides the actual value of 
QD upto an absolute error $\varepsilon\sim10^{-3}$.

The variations of QWD and CQWD, as functions of $b$, are depicted in Fig. \ref{24be1}(b) with the inset demonstrating the corresponding 
variations of $\varepsilon$. Similar to the case of QD, the optimal measurement observable is $\sigma^{z}$ for $b<0.22$, 
and $\sigma^{x}$ for $b\geq0.22$, which can be used to determine analytic expression of $W_c$ as
\begin{eqnarray}
 W_c=\min[\tilde{S}_1,\tilde{S}_2]+S(\rho_b),
\end{eqnarray}
where the functions $\tilde{S}_1$ and $\tilde{S}_2$ are given by 
\begin{eqnarray}
 \tilde{S}_1&=&\frac{1}{1+7b}\Big[1+b+(1+7b)\log_2(1+7b)-6b\log_2b\nonumber \\
          &&-\frac{1}{2}\sum_{i=1}^2\zeta_i\log_2\zeta_i\Big], \\
 \tilde{S}_2&=&-\frac{1}{2}\sum_{i,j=1}^{2}(\tau_{ij}\log_2\tau_{ij}+\tau^\prime_{ij}\log_{2}\tau^\prime_{ij}-\tau_{ij}-\tau_{ij}^\prime).\nonumber\\
\end{eqnarray}
The quantities $\zeta_i$, $\tau_{ij}$, and $\tau_{ij}^\prime$ are given in Eq. (\ref{quant}).
% \begin{eqnarray}
% \eta_i&=&1+b+(-1)^i\sqrt{1-b^2},\nonumber \\
% \kappa_{ij}&=&\frac{1}{8(1+7b)}\Big(1+9b+(-1)^i\sqrt{1-x^2}+(-1)^j\omega_{i}\Big),\nonumber \\
% \kappa^\prime_{ij}&=&\frac{1}{8(1+7b)}\Big(1+5b+(-1)^i\sqrt{1-x^2}+(-1)^j\omega^\prime_{i}\Big),\nonumber \\
% \end{eqnarray}
% with $\omega_i$ and $\omega^\prime_i$ being given in Eq. (\ref{omega}).
In the case of QWD, the point at which the sudden change takes place $(b=0.22)$ is different from that for 
QD $(b=0.15)$. Note that the behaviors of both constrained as well as actual QWD as functions of $b$ are similar to the behaviors
of the respective varieties of QD, and the VE, in both cases, also show similar variations. The maximum value of 
$\varepsilon$, in the case of QWD, is also of the order of $10^{-3}$. 

\begin{figure}
 \includegraphics[scale=0.675]{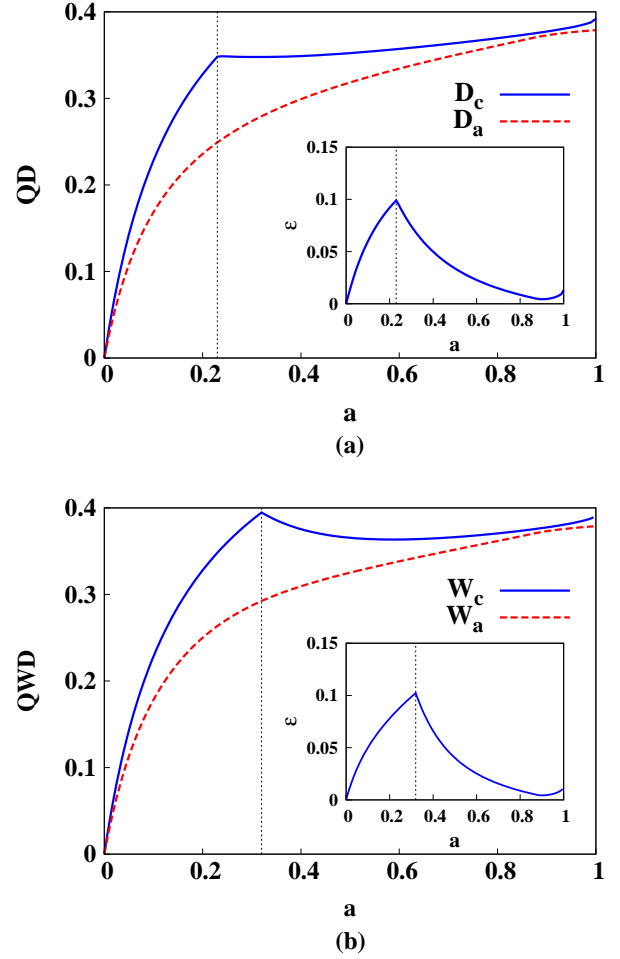}
 \caption{(Color online.) Variation of QD (a) and QWD (b) against $a$ in the case of the PPT BE state $\rho_{a}$ in 
 $\mathbb{C}^{3}\otimes\mathbb{C}^{3}$ systems. (Inset) Variation of the corresponding VE, $\varepsilon$, as a function of 
 $a$. The error is maximum at $a=0.23$ in the case of QD, and at $a=0.32$ in the case of QWD. 
 The quantities $D_c$, $D_a$, $W_c$, $W_a$, and $\varepsilon$ are in bits, while $a$ is dimensionless.}
 \label{33be1}
\end{figure}

\subsection{\texorpdfstring{$\mathbb{C}^{3}\otimes\mathbb{C}^{3}$}{} systems}
\label{3x3systems}

\subsubsection*{\textbf{Case 1}}

As an example of a BE state in a $\mathbb{C}^{3}\otimes\mathbb{C}^{3}$ systems, we first consider the following state \cite{be1}:
\begin{eqnarray}
 \rho_a =\frac{1}{8a + 1}
 \left(
 \begin{array}{ccccccccc}
  a & 0 & 0 & 0 & a & 0 & 0 & 0 & a \\
  0 & a & 0 & 0 & 0 & 0 & 0 & 0 & 0\\
  0 & 0 & a & 0 & 0 & 0 & 0 & 0 & 0 \\
  0 & 0 & 0 & a & 0 & 0 & 0 & 0 & 0\\
  a & 0 & 0 & 0 & a & 0 & 0 & 0 & a \\
  0 & 0 & 0 & 0 & 0 & a & 0 & 0 & 0 \\
  0 & 0 & 0 & 0 & 0 & 0 & f^{\prime}_{a}  & 0 & g^{\prime}_{a} \\
  0 & 0 & 0 & 0 & 0 & 0 & 0 & a & 0 \\
  a & 0 & 0 & 0 & a & 0 & g^{\prime}_{a} & 0 & f^{\prime}_{a}
 \end{array}
 \right),
 \label{3x3state1}
\end{eqnarray}
where $0\leq a\leq1$, and the functions $f^{\prime}_{a}=f_{b=a}$ and $g^{\prime}_{a}=g_{b=a}$, with $f_{b}$ and $g_{b}$ given by 
Eq. (\ref{fbgb}). Note that similar to the state, given in Eq. (\ref{2x4state}), 
the state is BE for the entire range of $a$. However, unlike the $\mathbb{C}^{2}\otimes\mathbb{C}^{4}$ state discussed in 
Sec. \ref{2x4system}, the local measurement must be performed on a subsystem. We consider the set  
$\{|0\rangle,|1\rangle,|2\rangle\}$ as the computational basis for each subsystem of dimension $d=3$. The triad, in this case,
consists of the measurements corresponding to the observables $S^{x}$, $S^{y}$, and $S^{z}$, where $S^{\beta}, \beta=x,y,z$, 
are the spin operators for a spin-$1$ particle. 
Fig. \ref{33be1}(a) demonstrates the variations of QD and CQD as functions of $a$ for the entire range $[0,1]$. 
Note that the VE (shown in the inset) is non-zero for the entire range of $a$, attaining its maximum at $a=0.23$. 
The optimization of $D_{c}$, is obtained for the measurement corresponding to the observable $S^{z}$ when 
$a<0.23$, while for $a\geq0.23$, the optimal measurement observable is $S^{x}$. Similar results are obtained in the 
case of QWD and CQWD (Fig. \ref{33be1}(b)) where the maximum 
VE occurs at $a=0.32$. In both the cases (QD and QWD), the maximum VE is of the order of $\sim10^{-1}$, which is much higher compared 
to the same in previous examples. 

\begin{figure}
\includegraphics[scale=0.675]{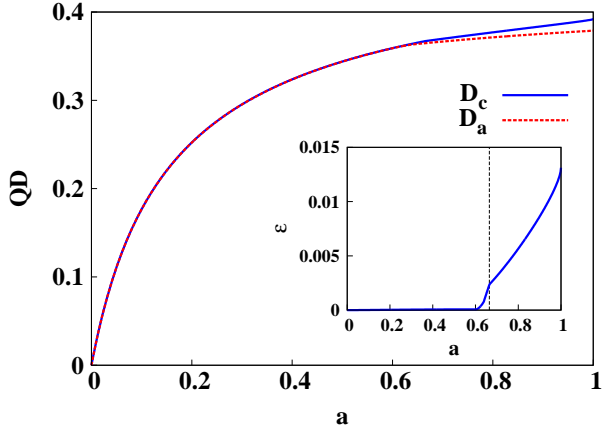}
\caption{(Color online.) Variation of $D_{c}$ and $D_{a}$ for $\rho_{a}$, 
as functions of the state parameter $a$, when measurement is performed over the subsystem $B$. 
(Inset) Variation of VE, $\varepsilon$, as a function of $a$. 
 The quantities $D_{c}$, $D_{a}$, and $\varepsilon$ are in bits, while $a$ is dimensionless.}
\label{bpart}
\end{figure}

An interesting observation comes from swapping the subsystem over which the local measurement is performed since the state $\rho_{a}$
is asymmetric over an exchange of the subsystems $A$ and $B$. If one optimizes QD of $\rho_{a}$ over a complete set of local 
measurements performed on the party $B$ instead of $A$, it is observed that the VE reduces drastically in comparison to 
the same obtained when measurement is performed over the party $A$. The variation of the corresponding $D_{c}$ and $D_{a}$
along with VE against the state parameter $a$ is given in Fig. \ref{bpart}. The VE attains a non-zero value (from the zero value) at 
$a=0.6$, while a sudden change in the variation profile is observed at $a=0.665$. This change is due to a transition of the optimal 
measurement observable from $S^{y}$ to $S^{x}$ at $a=0.665$. Note that the maximum value of VE in the region $a\geq0.6$ is $\sim10^{-2}$,
in contrast to the previous case.   
Note also that analytical expressions for CQD and CQWD can be obtained in a procedure similar as in the 
previous case.

\begin{figure}
 \includegraphics[scale=0.675]{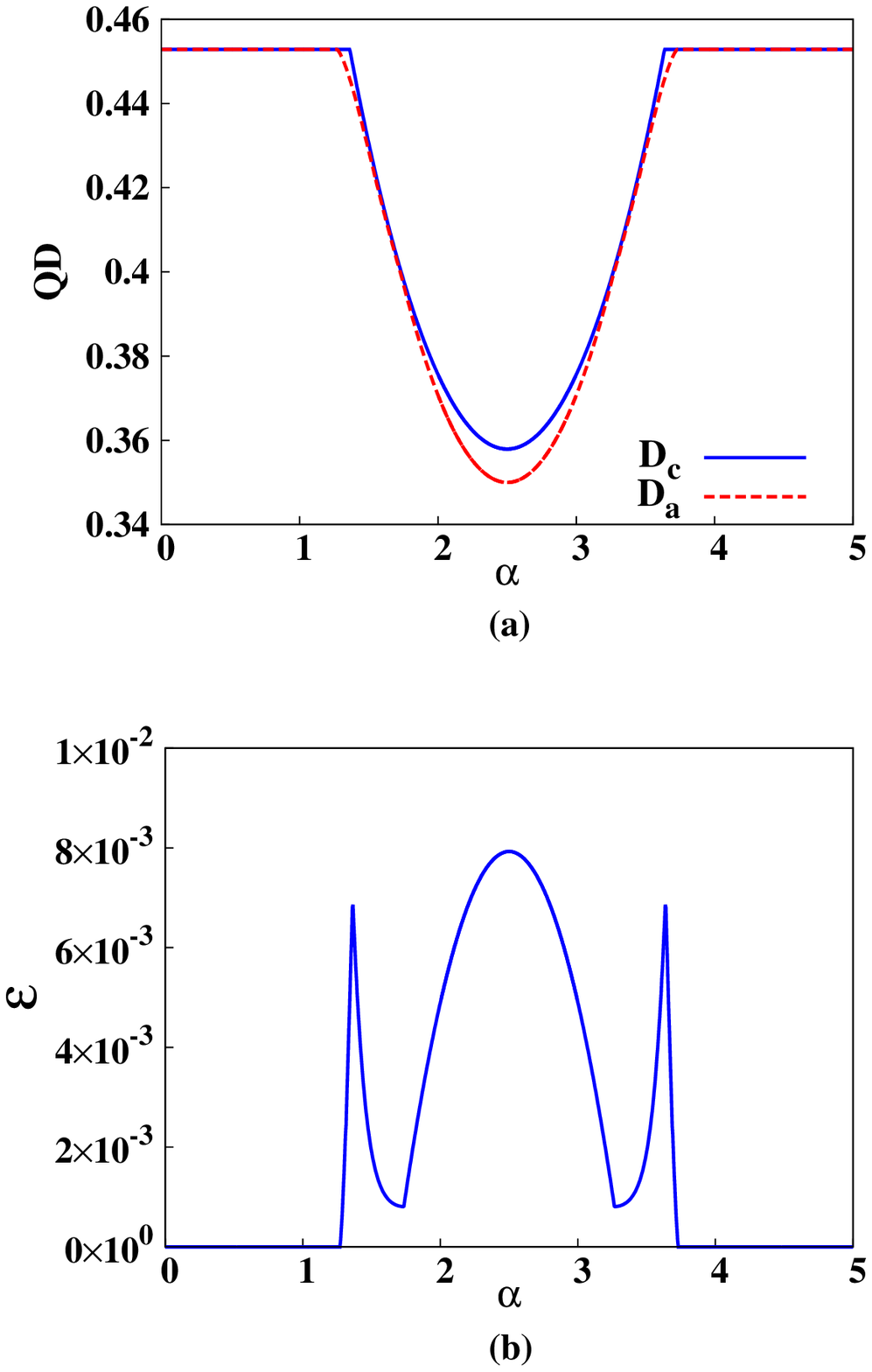}
 \caption{(Color online.) Variation of QD (a) and the corresponding VE (b) as functions of $\alpha$ in the case of the PPT BE state 
 $\varrho_{\alpha}$, given in Eq. (\ref{3x3state2}).
 The quantities $D_{c}$, $D_{a}$, and $\varepsilon$ are in bits, while $\alpha$ is dimensionless.}
 \label{33be2}
\end{figure}

\subsubsection*{\textbf{Case 2}}

Let us take another example of a PPT BE state in $\mathbb{C}^{3}\otimes\mathbb{C}^{3}$, given by \cite{be3}:
\begin{equation}
 \varrho_{\alpha} = \frac{2}{7}|\psi\rangle\langle \psi| + \frac{\alpha}{7}\varrho_+ + \frac{5 -\alpha}{7}\varrho_-,
 \label{3x3state2}
\end{equation}
where $ \varrho_+=(|01\rangle\langle 01|+|12\rangle\langle 12|+|20\rangle\langle 20|)/3$, 
$\varrho_-=(|10\rangle\langle 10| + |21\rangle\langle 21| + |02\rangle\langle 02|)/3$, 
$|\psi\rangle = \frac{1}{\sqrt{3}}\sum_{i = 0}^2 |ii\rangle$, and $0\leq\alpha\leq5$.  The state 
is separable for $2\leq\alpha\leq3$, BE for $3<\alpha\leq4$ and $1<\alpha\leq2$, while distillable for $4<\alpha\leq5$ and 
$0<\alpha\leq1$ \cite{be3}. Fig. \ref{33be2}(a) depicts the variation of QD and CQD as functions of the state parameter 
$\alpha$ over its entire range. Note that the QD as well as the CQD remains constant over the range of $\alpha$ in which entanglement 
is distillable, whereas both of them attains a minimum at $\alpha=2.5$ in the separable region. In the BE region, $1<\alpha\leq2$, 
the value of QD as well as CQD remains constant up to $\alpha=1.36$, and then decreases for increasing $\alpha$ in the region 
$1.36\leq\alpha\leq2$. Besides, in the BE region, $3<\alpha\leq4$, the value of QD as well as CQD increases with increasing $\alpha$ 
up to $\alpha=3.64$, and then becomes constant. The corresponding $\varepsilon$ is plotted against $\alpha$ in 
Fig. \ref{33be2}(b). Note that the maximum VE is committed in the 
separable region (of the order of $\sim10^{-2}$), whereas it is smaller (of the order of $\sim10^{-3}$) in the BE region. Clearly, the 
maximum VE is relatively higher in the present case in comparison to the $\mathbb{C}^{2}\otimes\mathbb{C}^{4}$ BE state, 
$\rho_{b}$ (Eq. (\ref{2x4state})), but considerably lower when compared to the $\mathbb{C}^{3}\otimes\mathbb{C}^{3}$ BE state, 
$\rho_{a}$ (Eq. (\ref{3x3state1})). When QD is constant, the optimal measurement observable is $S^{z}$ while it changes to 
either $S^{x}$ or $S^{y}$ when value of QD increases or decreases. Note that in the latter case, choice of $S^{x}$ or $S^{y}$ as the 
optimal measurement observable are equivalent in the context of optimizing QD. Analytic
expression for $D_c$ can be obtained using the above analysis, as shown in the previous cases.

We conclude the discussion on the state $\varrho_{\alpha}$ by pointing out that the QWD of the state coincides with the QD since 
$\varrho_{\alpha}$ has maximally mixed marginals, i.e., the local density operator 
of each of the parties is proportional to identity in $\mathbb{C}^{3}$ \cite{wdef_group}.

\section{Concluding Remarks}
\label{conclude}

To summarize, we have addressed the question whether the computational complexity of information theoretic measures of 
quantum correlations such as quantum discord and quantum work deficit can be reduced by performing the optimization involved
over a constrained subset of local projectors instead of the complete set. We have considered four plausible constructions 
of such a restricted set, and shown that the average absolute error, in the case of two-qubit mixed states with different ranks,  
dies down fast with the increase in the size of the set. 
Quantitative investigation of the reduction of error with the increase in the size of the restricted set has been performed 
with a comparative study between quantum discord and quantum work deficit, and the
corresponding scaling exponents have been estimated. We have also considered a general two-qubit state up to local unitary 
transformation, and have shown that the computation of measures like quantum discord and quantum work deficit can be 
made considerably easier by carefully choosing the restricted set of projectors. We have also pointed out that insight about 
constructing the constrained set can be gathered from the probability distribution of the optimizing parameters.

If a very special restricted set consisting of the 
% Forming a very
% special restricted set consisting of the 
projection measurements corresponding to only the three Pauli matrices is considered, 
we have demonstrated that the state space of a two-qubit system 
contains a large fraction of states for which exact minimizations of quantum discord and quantum work deficit are 
obtained only on this set, resulting vanishing error.
% possible by performing 
% minimization over this set only.
We have also pointed out that this feature can be utilized to obtain closed-form expressions of quantum 
correlation measures up to small error for some special classes of states, which can be used to study physical systems 
such as quantum spin models. The usefulness of this methodology has been demonstrated in the finite size scaling analysis 
of the well-known transverse-field $XY$ model at zero temperature, and for the thermal state of a two-qubit $XY$ model 
in an external inhomogeneous transverse field. 
Moreover, we have found that the absolute error in the value of 
quantum correlation calculated using the constrained set in the case of two-qubit PPT as well as PPT BE states is low compared 
to the NPPT states. 
% The results found in this study are expected 
% to be useful in the computation of quantum correlation measures like quantum discord and quantum work deficit in scenarios 
% where a restriction over the local projective measurement is active. 
The investigations show that these measures can be obtained with high accuracy even when restrictions in the optimizations 
involved in their definitions are employed, thereby reducing the computational difficulties in their evaluation. Although we 
have investigated only information theoretic measures, this study gives rise to a possibility to overcome challenges in the 
computation of other quantum correlation measures that involve optimization over some set.


\begin{thebibliography}{100}
\bibitem{ent_horodecki}       R. Horodecki, P. Horodecki, M. Horodecki, and K. Horodecki, Rev. Mod. Phys. {\bf81}, 865 (2009).
\bibitem{ent_th}              C. H. Bennett and S.J. Wiesner, Phys. Rev. Lett. {\bf 69}, 2881 (1992);
                              C. H. Bennett, G. Brassard, C. Cr\'epeau, R. Jozsa, A. Peres, and W. K. Wootters, Phys. Rev. Lett. {\bf 70}, 1895 (1993);
                              R. Raussendorf and H. J. Briegel, Phys. Rev. Lett. {\bf 86}, 5188 (2001); 
                              P. Walther, K. J. Resch, T. Rudolph, E. Schenck, H. Weinfurter, V. Vedral, M. Aspelmeyer, and A. Zeilinger, Nature {\bf 434}, 169 (2005); 
                              H. J. Briegel, D. Browne, W. D\"ur, R. Raussendorf,  and M. van den Nest, Nat. Phys. {\bf 5}, 19 (2009).
 \bibitem{ent_exp}            J. M. Raimond, M. Brune, S. Haroche, Rev. Mod. Phys. \textbf{73}, 565 (2001);
                              D. Leibfried, R. Blatt, C. Monroe, and D. Wineland, Rev. Mod. Phys. \textbf{75}, 281 (2003);
                              L. M. K. Vandersypen, I.L Chuang, Rev. Mod. Phys. \textbf{76}, 1037 (2005); 
                              K. Singer, U. Poschinger, M. Murphy, P. Ivanov, F. Ziesel, T. Calarco, F. Schmidt-Kaler,  Rev. Mod. Phys. \textbf{82}, 2609 (2010);
                              H. Haffner, C. F. Roose, R. Blatt, Phys. Rep. \textbf{469}, 155 (2008);
                              L.-M. Duan, C. Monroe, Rev. Mod. Phys. \textbf{82}, 1209 (2010);
                              J.-W. Pan, Z.-B. Chen, C.-Y. Lu, H. Weinfurter, A. Zeilinger, M. \.Zukowski, Rev. Mod. Phys. \textbf{84}, 777 (2012).
 \bibitem{distil}             C. H. Bennett, G. Brassard, S. Popescu, B. Schumacher, J. A. Smolin, and W. K. Wootters, Phys. Rev. Lett. \textbf{76}, 722 (1996); 
                              M. Horodecki, P. Horodecki, and R. Horodecki, Phys. Rev. Lett. \textbf{78}, 574 (1997);                    
                              M. Murao, M. B. Plenio, S. Popescu, V. Vedral, and P. L. Knight, Phys. Rev. A \textbf{57}, R4075 (1998);                    
                              W. D\"{u}r, J. I. Cirac, and R. Tarrach, Phys. Rev. Lett. \textbf{83}, 3562 (1999);
                              M. Horodecki and P. Horodecki, Phys. Rev. A \textbf{59}, 4206 (1999);
                              W. D\"{u}r and J. I. Cirac, Phys. Rev. A \textbf{61}, 042314 (2000);                    
                              W. D\"{u}r, J. I. Cirac, M. Lewenstein, and D. Bru{\ss}, Phys. Rev. A \textbf{61}, 062313 (2000);
                              G. Alber, A. Delgado, N. Gisin, and I. Jex, J. Phys. A. \textbf{34}, 8821 (2001);                    
                              J. Dehaene, M. Van den Nest, B. D. Moor, and F. Vestraete, Phys. Rev. A \textbf{67}, 022310 (2003);
                              W. D\"{u}r, H. Aschaur, and H. J. Briegel, Phys. Rev. Lett. \textbf{91}, 107903 (2003);
                              E. Hostens, J. Dehaene, and B. D. Moor, arXiv:quant-ph/0406017 (2004);
                              K. G. H. Vollbrecht and F. Vestraete, Phys. Rev. A \textbf{71}, 062325 (2005);                    
                              A. Miyake and H. J. Briegel, Phys. Rev. Lett. \textbf{95}, 220501 (2005);                    
                              H. Aschaur, W. D\"{u}r, and H. J. Briegel, Phys. Rev. A \textbf{71}, 012319 (2005);
                              E. Hostens, J. Dehaene, and B. D. Moor, Phys. Rev. A \textbf{73}, 042316 (2006);
                              E. Hostens, J. Dehaene, and B. D. Moor, Phys. Rev. A \textbf{74}, 062318 (2006);                    
                              A. Kay, J. K. Pachos, W. D\"{u}r, and H. J. Briegel, New. J. Phys. \textbf{8}, 147 (2006);
                              C. Kruszynska, A. Miyake, H. J. Briegel, and W. D\"{u}r, Phys. Rev. A \textbf{74}, 052316 (2006);
                              S. Glancy, E. Knill, and H. M. Vasconcelos, Phys. Rev. A \textbf{74}, 032319 (2006).  
 \bibitem{nl_w_ent}           C. H. Bennett, D. P. DiVincenzo, C. A. Fuchs, T. Mor, E. Rains, P. W. Shor, J. A. Smolin, and W. K. Wootters, Phys. Rev. A {\bf 59}, 1070 (1999); 
                              C. H. Bennett, D.P. DiVincenzo, T. Mor, P.W. Shor, J.A. Smolin, and B.M. Terhal, Phys. Rev. Lett. {\bf 82}, 5385 (1999); 
                              D. P. DiVincenzo, T. Mor, P. W. Shor, J. A. Smolin, and B. M. Terhal Commun. Math. Phys. {\bf 238}, 379 (2003).
 \bibitem{opt_det}            A. Peres and W.K. Wootters, Phys. Rev. Lett. \textbf{66}, 1119 (1991); 
                              J. Walgate, A.J. Short, L. Hardy, and V. Vedral,ibid. \textbf{85}, 4972 (2000); 
                              S. Virmani, M.F. Sacchi, M.B. Plenio, and D. Markham, Phys. Lett. A \textbf{288}, 62 (2001); 
                              Y.-X. Chen and D. Yang, Phys. Rev. A 64, 064303 (2001);ibid. \textbf{65}, 022320 (2002); 
                              J. Walgate and L. Hardy, Phys. Rev. Lett. \textbf{89}, 147901 (2002); 
                              M. Horodecki, A. Sen(De), U. Sen, and K. Horodecki, ibid. \textbf{90}, 047902 (2003); 
                              W. K. Wootters, Int. J. Quantum Inf. \textbf{4}, 219 (2006).
 \bibitem{dqci}               E. Knill and R. Laflamme, Phys. Rev. Lett. {\bf81}, 5672 (1998);
                              A. Datta, A. Shaji, and C. M. Caves, Phys. Rev. Lett. \textbf{100}, 050502 (2008);
                              B. P. Lanyon, M. Barbieri, M. P. Almeida, and A. G. White, Phys. Rev. Lett. 101, 200501 (2008).
 \bibitem{dakic_geom}         B. Daki\'{c}, V. Vedral, and \~{C}. Brukner, Phys. Rev. Lett. \textbf{105}, 190502 (2010).
 \bibitem{disc_group}         L. Henderson and V. Vedral, J. Phys. A \textbf{34}, 6899 (2001);
                              H. Olivier and W. H. Zurek, Phys. Rev. Lett. \textbf{88}, 017901 (2001);
                              W. H. Zurek, Rev. Mod. Phys. \textbf{75}, 715 (2003).                              
 \bibitem{wdef_group}         J. Oppenheim, M. Horodecki, P. Horodecki and R. Horodecki, Phys. Rev. Lett. \textbf{89}, 180402 (2002);
                              M. Horodecki, K. Horodecki, P. Horodecki, R. Horodecki, J. Oppenheim, A. Sen(De), and U. Sen, Phys. Rev. Lett. \textbf{90}, 100402 (2003); 
                              I. Devetak, Phys. Rev. A \textbf{71}, 062303 (2005);
                              M. Horodecki, P. Horodecki, R. Horodecki, J. Oppenheim, A. Sen(De), U. Sen, and B. Synak-Radtke, Phys. Rev. A \textbf{71}, 062307 (2005). 
 \bibitem{celeri_2011}        L. C. C\'{e}leri, J. Maziero, and R. M. Serra, Int. J. Quant. Inf. \textbf{9}, 1837 (2011), and the references therein. 
 \bibitem{modi_rmp_2012}      K. Modi, A. Brodutch, H. Cable, T. Paterek, and V. Vedral, Rev. Mod. Phys. \textbf{84}, 1655 (2012), and the references 
                              therein.                            
 \bibitem{ent_disc}           M. Koashi, and A. Winter, Phys. Rev. A \textbf{69}, 022309 (2004); 
                              F. F. Fanchini, M. F. Cornelio, M. C. de Oliveira, and A. O. Caldeira, Phys. Rev. A \textbf{84}, 012313 (2011); 
                              S. Yu, C. Zhang, Q. Chen, and C. Oh, arXiv:1102.1301 [quant-ph] (2011); 
                              C. Zhang, S. Yu, Q. Chen, and C. Oh, Phys. Rev. A \textbf{84}, 052112 (2011); 
                              L.-X. Cen, X.-Q. Li, J. Shao, and Y. Yan, Phys. Rev. A \textbf{83}, 054101 (2011); 
                              F. Galve, G. Giorgi, and R. Zambrini, Europhys. Lett. \textbf{96}, 40005 (2011); 
                              A. Al-Qasimi and D. F. V. James, Phys. Rev. A \textbf{83}, 032101 (2011); 
                              T. K. Chuan, J. Maillard, K. Modi, T. Paterek, M. Paternostro, and M. Piani, Phys. Rev. Lett. \textbf{109}, 070501 (2012); 
                              F. Lastra, C. Lopez, L. Roa, and J. Retamal, Phys. Rev. A \textbf{85}, 022320 (2012); 
                              F. F. Fanchini, L. K. Castelano, M. F. Cornelio, and M. C. de Oliveira, New J. Phys. \textbf{14}, 013027 (2012); 
                              Z. Xi, X.-M. Lu, X. Wang, and Y. Li, Phys. Rev. A \textbf{85}, 032109 (2012).  
 \bibitem{distil_disc}        A. Streltsov, H. Kampermann, and D. Bru{\ss}, Phys. Rev. Lett. \textbf{106}, 160401 (2011); 
                              M. Piani, S. Gharibian, G. Adesso, J. Calsamiglia, P. Horodecki, and A. Winter, Phys. Rev. Lett. \textbf{106}, 220403 (2011);
                              M. F. Cornelio, M. C. de Oliveira, and F. F. Fanchini, Phys. Rev. Lett. \textbf{107}, 020502 (2011); 
                              X. Q. Yan, G. H. Liu, and J. Chee, Phys. Rev. A \textbf{87}, 022340 (2013).      
 \bibitem{be1}                P. Horodecki, Phys. Lett. A \textbf{232}, 333 (1997).
 \bibitem{be2}                M. Horodecki, P. Horodecki, and R. Horodecki, Phys. Rev. Lett. \textbf{80}, 5239 (1998).
 \bibitem{be3}                P. Horodecki, M. Horodecki, and R. Horodecki, Phys. Rev. Lett. \textbf{82}, 1056, (1999).
 \bibitem{be4}                D. Bruss, A. Peres, Phys. Rev. A. \textbf{61}, 030301(R) (2000). 
 \bibitem{be5}                W. D\"{u}r and J. I. Cirac, Phys. Rev. A \textbf{62}, 022302 (2000);
                              P. Horodecki, J. A. Smolin, B. M. Terhal, and A. V. Thapliyal, Theor. Comput. Sc. \textbf{292}, 589 (2003);
                              J. Lavoie, R. Kaltenbaek, M. Piani, and  K. J. Resch, Nature Physics \textbf{6}, 827 (2010);               
                              J. Lavoie, R. Kaltenbaek, M. Piani, and  K. J. Resch, Phys. Rev. Lett. \textbf{105}, 130501 (2010);
                              J. DiGuglielmo, A. Samblowski, B. Hage, C. Pineda, J. Eisert, and R. Schnabel, Phys. Rev. Lett. \textbf{107}, 240503 (2011); 
                              E. Amselem, M. Sadiq, and M. Bourennane, Scientific Reports \textbf{3}, 1966 (2013). 
 \bibitem{be6}                T. Moroder, O. Gittsovich, M. Huber, and O. G\"{u}hne, Phys. Rev. Lett. \textbf{113}, 050404 (2014).
 \bibitem{npptbe}             D. P. DiVincenzo, P. W. Shor, J. A. Smolin, B. M. Terhal, and A. V. Thapliyal, Phys. Rev. A \textbf{61}, 062312 (2000);
                              B. Kraus, M. Lewenstein, and J. I. Cirac, Phys. Rev. A \textbf{65}, 042327 (2002);
                              J. Watrous, Phys. Rev. Lett. \textbf{93}, 010502 (2004). 
 \bibitem{luo_pra_2008}       S. Luo, Phys. Rev. A \textbf{77}, 042303 (2008).
 \bibitem{disc_2q}            M. Ali, A. R. P. Rau, and G. Alber, Phys. Rev. A \textbf{81}, 042105 (2010); 
                              \textit{ibid.} \textbf{82}, 069902(E), (2010);
                              X.-M. Lu, J. Ma, Z. Xi, and X. Wang, Phys. Rev. A \textbf{83}, 012327 (2011);
                              D. Girolami and G. Adesso, Phys. Rev. A \textbf{83}, 052108 (2011);
                              Q. Chen, C. Zhang, S. Yu, X. X. Yi, and C. H. Oh, Phys. Rev. A \textbf{84}, 042313 (2011).
 \bibitem{disc_mq_hd}         M. Ali, J. Phys. A: Math. Theor. \textbf{43}, 495303 (2010);
                              S. Vinjanampathy and A. R. P. Rau, J. Phys. A: Math. Theor. \textbf{45}, 095303 (2012); 
                              B. Ye, Y. Liu, J. Chen, X. Liu, and Z. Zhang, Quant. Inf. Process. \textbf{12}, 2355 (2013).
 \bibitem{huang_disc_np}      Y. Huang, New. J. Phys. \textbf{16} (3), 033027 (2014).                             
 \bibitem{disc_expt}          M. A. Yurishchev, Phys. Rev. B \textbf{84}, 024418 (2011); 
                              R. Auccaise, J. Maziero, L. C. C ́eleri, D. O. Soares-Pinto, E. R. deAzevedo, T. J. Bonagamba, R. S. Sarthour, 
                              I. S. Oliveira, and R. M. Serra, Phys. Rev. Lett. \textbf{107}, 070501 (2011);
                              G. Passante, O. Moussa, D. A. Trottier, and R. Laflamme, Phys. Rev. A \textbf{84}, 044302 (2011);
                              L. S. Madsen, A. Berni, M. Lassen, and U. L. Andersen, Phys. Rev. Lett. \textbf{109}, 030402 (2012).
                              M. Gu, H. M. Chrzanowski, S. M. Assad, T. Symul, K. Modi, T. C. Ralph, V. Vedral, 
                              and P. K. Lam, Nature Phys. 8, \textbf{671} (2012);
                              R. Blandino, M. G. Genoni, J. Etesse, M. Barbieri, M. G. A. Paris, P. Grangier, 
                              and R. Tualle-Brouri, Phys. Rev. Lett. \textbf{109}, 180402 (2012);
                              U. Vogl, R. T. Glasser, Q. Glorieux, J. B. Clark, N. V. Corzo, and P. D. Lett, Phys. Rev. A \textbf{87}, 010101(R) (2013);
                              C. Benedetti, A. P. Shurupov, M. G. A. Paris, G. Brida, and M. Genovese, Phys. Rev. A \textbf{87}, 052136 (2013).
 \bibitem{num-err_group}      Y. Huang, Phys. Rev. A \textbf{88}, 014302 (2013);
                              M. Namkung, J. Chang, J. Shin, and Y. Kwon, Int. J. Theor. Phys. \textbf{54}, 3340 (2015).                        
 \bibitem{disc_num_anal}      F. F. Fanchini, T. Werlang, C. A. Brasil, L. G. E. Arruda, and A. O. Caldeira, Phys. Rev. A \textbf{81}, 052107 (2010);
                              B. Li, Z. -X. Wang, and S. -M. Fei, Phys. Rev. A \textbf{83}, 022321 (2011).
\bibitem{freezing_paper}     T. Chanda, A. K. Pal, A. Biswas, A. Sen(De), and U. Sen, Phys. Rev. A \textbf{91}, 062119 (2015).                              
%  \bibitem{state_def}          J. Schlienz and G. Mahler, Phys. Rev. A \textbf{52}, 4396 (1995);
%                               F. Verstraete, J. Dehaene, and B. DeMoor, Phys. Rev. A \textbf{64}, 010101(R) (2001).
 \bibitem{qet}                C. W. Helstrom, Quantum Detection and Estimation Theory (Academic Press, New York, 1976); 
                              A.S. Holevo, Statistical Structure of Quantum Theory, Lect. Not. Phys. 61 (Springer, Berlin, 2001);
                              C. W. Helstrom, Phys. Lett. A \textbf{25}, 1012 (1967);
                              H. P. Yuen, M. Lax, IEEE Trans. Inf. Th. \textbf{19}, 740 (1973);
                              C. W. Helstrom, R. S. Kennedy, IEEE Trans. Inf. Th. \textbf{20}, 16 (1974);
                              S. Braunstein and C. Caves, Phys. Rev. Lett. \textbf{72}, 3439 (1994);
                              S. Braunstein, C. Caves, and G. Milburn, Ann. Phys. \textbf{247}, 135 (1996);
                              M. G. A. Paris,  Int. J. Quant. Inform., \textbf{07}, 125 (2009), and the references therein.   
 \bibitem{ppt-ph}             A. Peres, Phys. Rev. Lett. \textbf{77}, 1413 (1996);
                              M. Horodecki, P. Horodecki, and R. Horodecki, Phys. Lett. A \textbf{223}, 1 (1996).
 \bibitem{xstate}             T. Eu and J. H. Eberly, Quant. Inform. Comput. \textbf{7}, 459 (2007).                             
 \bibitem{xy_group}           E. Lieb, T. Schultz, and D. Mattis, Ann. Phys. \textbf{16}, 407 (1961);
                              E. Barouch, B.M. McCoy, and M. Dresden, Phys. Rev. \textbf{2}, 1075 (1970); 
                              P. Pfeuty, Ann. Phys. \textbf{57}, 79 (1970);                              
                              E. Barouch and B.M. McCoy, Phys. Rev. \textbf{3}, 786 (1971).   
 \bibitem{tot_corr}           N. J. Cerf, C. Adami, Phys. Rev. Lett. \textbf{79}, 5194 (1997);
                              B. Groisman, S. Popescu, and A. Winter, Phys. Rev. A \textbf{72}, 032317 (2005).                                    
 \bibitem{qd_geom}            S. Luo and S. Fu, Phys. Rev. A \textbf{82}, 034302 (2010); 
                              X. -M. Lu, Z. -J. Xi, Z. Sun, and X. Wang, Quant. Info. Comput., \textbf{10}, 0994 (2010); 
                              T. Debarba, T. O. Maciel, and R. O. Vianna, Phys. Rev. A \textbf{86}, 024302 (2012); 
                              J. -S. Jin, F. -Y Zhang, C.-S. Yu, and H. -S. Song, J. Phys. A: Math. Theor. \textbf{45}, 115308 (2012); 
                              J. D. Montealegre, F. M. Paula, A. Saguia, and M. S. Sarandy, Phys. Rev. A \textbf{87}, 042115 (2013); 
                              D. Spehner and M. Orszag, New J. Phys. \textbf{15}, 103001 (2013); 
                              D. Spehner and M. Orszag, J. Phys. A: Math. Theor. \textbf{47}, 035302 (2014).
 \bibitem{qd_geom_adesso}     D. Girolami and G. Adesso, Phys. Rev. A \textbf{84}, 052110 (2011).                             
 \bibitem{qd_geom_use}        B. Daki\'{c}, Y. O. Lipp, X. Ma, M. Ringbauer, S. Kropatschek, S. Barz, T. Paterek, V. Vedral, A. Zeilinger, 
                              \u{C}aslav Brukner, and P. Walther, Nature Phys. \textbf{8}, 666 (2012).                              
 \bibitem{qd_geom_prob}       M. Piani, Phys. Rev. A \textbf{86}, 034101 (2012).
 \bibitem{qd_geom_one}        F. M. Paula, Thiago R. de Oliveira, and M. S. Sarandy, Phys. Rev. A \textbf{87}, 064101 (2013).
 \bibitem{bell_x_onenorm}     F. Ciccarello, T. Tufarelli, and V Giovannetti, New J. Phys. \textbf{16}, 013038 (2014).                            
 \bibitem{op_int}             L. Roa, J. C. Retamal, and M. Alid-Vaccarezza, Phys. Rev. Lett. \textbf{107}, 080401 (2011);
                              V. Madhok and A. Datta, Phys. Rev. A \textbf{83}, 032323 (2011);
                              V. Madhok and A. Datta, Int. J. Mod. Phys. B \textbf{27}, 1345041 (2013).
 \bibitem{r2ppt}              A two-qubit rank-$2$ state can be written as 
                              $\rho=p|\psi\rangle\langle\psi|+(1-p)|\phi\rangle\langle\phi|$,
                              where $0\leq p\leq1$, and $|\psi\rangle$ and $|\phi\rangle$ are mutually orthogonal. The state 
                              $|\psi\rangle$, in Schmidt decomposed form, can be written as 
                              $|\psi\rangle=\alpha|00\rangle+\beta|11\rangle$, where $\alpha$, $\beta$, are real and 
                              $\alpha^{2}+\beta^2=1$. The partial transposition, $\rho^{T}$, of $\rho$ will always have a 
                              negative eigenvalue given by $\lambda=-|2p\alpha\beta-1|$. The state is PPT only when 
                              $\lambda=0$, which corresponds to a line in the state space of all rank-$2$ states. Therefore, 
                              the PPT states in the space of all rank-$2$ states form a set of measure zero.                                 
 \bibitem{volume}             K. \.{Z}yczkowski, P. Horodecki, A. Sanpera, and M. Lewenstein, Phys. Rev. A \textbf{58}, 883 (1998),
                              and references thereto. 
 \bibitem{xxz_group}          C. N. Yang and C. P. Yang, Phys. Rev. \textbf{150}, 321 (1966); 
                              C. N. Yang and C. P. Yang, Phys. Rev. \textbf{150}, 327 (1966);
                              A. Langari, Phys. Rev. B \textbf{58}, 14467 (1998);
                              D. V. Dmitriev, V. Y. Krivnov, A. A. Ovchinnikov, and A. Langari, 
                              J. Exp. Theor. Phys. \textbf{95}, 538 (2002);
                              S Mahdavifar, J. Phys.: Condens. Matter \textbf{19} 406222 (2007).                             
 \bibitem{rev_xy}             L. Amico, R. Fazio, A. Osterloh, and V. Vedral, Rev. Mod. Phys. \textbf{80}, 517 (2008); 
                              J. I. Latorre, and A. Rierra, J. Phys. A \textbf{42}, 504002 (2009); 
                              J. Eisert, M. Cramer, and M. B. Plenio, Rev. Mod. Phys. \textbf{82}, 277, (2010).   
 \bibitem{ent_xy_bp}          A. Osterloh, L. Amico, G. Falci, and R. Fazio, Nature {\bf 416}, 608 (2002); 
                              T. Osborne, and M. Nielsen, Phys. Rev. A {\bf 66}, 032110 (2002); 
                              G. Vidal, J. Latorre, E. Rico, and A. Kitaev, Phys.Rev. Lett. {\bf 90}, 227902 (2003).                              
 \bibitem{xy_discord}         R. Dillenschneider, Phys. Rev. B \textbf{78}, 224413 (2008);
                              M. S. Sarandy, Phys. Rev. A \textbf{80}, 022108 (2009).                              
 \bibitem{solid_xy}           M. Schechter and P. C. E. Stamp, Phys. Rev. B \textbf{78}, 054438 (2008).
 \bibitem{num_trap_xy}        X. -L. Deng, D. Porras, and J. I. Cirac, Phys. Rev. A \textbf{72}, 063407 (2005).                              
 \bibitem{trap_rev}           M. Lewenstein, A. Sanpera, V. Ahufinger, B. Damski, A.Sen(De), and U. Sen, Adv. Phys. \textbf{56}, 243 (2007).                             
 \bibitem{lab_xy}             R. Islam, E. E. Edwards, K. Kim, S. Korenblit, C. Noh, H. Carmichael, G.-D. Lin, L.-M. Duan, C.-C. Joseph Wang, J. K. Freericks, and C. Monroe,  Nature Commun. {\bf2}, 377 (2011);              
                              J. Struck, M. Weinberg, C. \"{O}lschläger, P. Windpassinger, J. Simonet, K. Sengstock, R. H\"{o}ppner, P. Hauke, A. Eckardt, M. Lewenstein, and  L. Mathey, Nature Physics \textbf{9}, 738 (2013), 
                              and references therein.   
 \bibitem{qpt_book}           B. K. Chakrabarti, A. Dutta, and P. Sen, Quantum Ising Phases and Transitions in Transverse Ising Models (Springer, Heidelberg, 1996); 
                             S. Sachdev, Quantum Phase Transitions (Cambridge University Press, Cambridge, 2011);
                             S. Suzuki, J. -I. Inou, B. K. Chakrabarti, Quantum Ising Phases and Transitions in Transverse Ising Models (Springer, Heidelberg, 2013).                                                            
\end{thebibliography}
\end{document}